\documentclass[final, 11pt, sort&compress, number, times, 5p]{elsarticle}
\usepackage{multirow}
\usepackage[colorlinks,pdftex,pdfencoding=auto]{hyperref}

\usepackage{listings}
\usepackage{tcolorbox}
\tcbuselibrary{listings,skins,breakable}
\newtcblisting{pythoncode}{
    listing only,
    listing options={
        language=Python,
        basicstyle=\ttfamily\small,
        keywordstyle=\color{blue}\bfseries,
        commentstyle=\color{gray}\itshape,
        stringstyle=\color{teal},
        breaklines=true,
        breakatwhitespace=true,
        columns=fullflexible
    },
    colback=gray!5,
    colframe=gray!80,
    boxrule=0.5mm,
    arc=2mm,
    left=2mm,
    right=2mm,
    top=1mm,
    bottom=1mm
}
\usepackage{graphicx}
\usepackage{amsmath}
\usepackage{amssymb}
\usepackage{braket}
\usepackage{xcolor}
\usepackage{float}

\usepackage{xurl}

\newcounter{bla}

\newcommand{\rtt}[2]{\href{https://seemps.readthedocs.io/en/latest/#1.html}{\texttt{#2}}}
\newcommand{\jj}[1]{#1}
\newcommand{\jg}[1]{#1}
\newcommand{\jr}[1]{#1}

\journal{Computer Physics Communications}

\begin{document}

\begin{frontmatter}



\title{SeeMPS: A Python-based Matrix Product State and Tensor Train Library}


\author[a,b]{Paula García-Molina}
\author[a]{Juan José Rodríguez-Aldavero}
\author[a,c]{Jorge Gidi}
\author[a]{Juan José García-Ripoll\corref{author}}

\cortext[author] {Corresponding author.\\\textit{E-mail address:} jj.garcia.ripoll@csic.es}
\address[a]{Instituto de Física Fundamental IFF-CSIC, Calle Serrano 113b, Madrid 28006}
\address[b]{Centre de Visió per Computador (CVC), Barcelona, Spain}

\address[c]{Millennium Institute for Research in Optics and Departamento de Física, Facultad de Ciencias Físicas y Matemáticas, Universidad de Concepción, Casilla 160-C, Concepción, Chile}

\begin{abstract}
We introduce SeeMPS, a Python library dedicated to implementing tensor network algorithms based on the well-known \jj{Matrix Product State (MPS), or Tensor Train (TT), formalism, together with its quantized version (QTT) for the representation of functions}. SeeMPS is implemented as a complete finite precision linear algebra package where exponentially large vector spaces are compressed using the MPS/TT formalism. It enables both low-level operations, such as vector addition, linear transformations, and Hadamard products, as well as high-level algorithms, including the \jj{approximate solution} of linear equations, eigenvalue computations, and \jj{Fourier transforms that, for suitably structured states, achieve an exponential speedup over the fast Fourier transform (FFT)}. This library can be used for traditional quantum many-body physics applications and also for quantum-inspired numerical analysis problems, such as solving \jj{partial differential equations (PDEs)}, interpolating\jj{,} and integrating multidimensional functions\jj{, and} sampling multivariate probability distributions.

\noindent \textbf{PROGRAM SUMMARY}
\begin{small}
\noindent
{\em Program Title:} SElf-Explaining Matrix-Product-State (SeeMPS) Library\\
{\em CPC Library link to program files:} (to be added by Technical Editor) \\
{\em Developer's repository link:} \url{https://github.com/juanjosegarciaripoll/seemps2} \\
{\em Licensing provisions (please choose one):} MIT  \\
{\em Programming language:}  Python, Cython\\
{\em Supplementary material:}  \url{https://seemps.readthedocs.io/}                           \\
{\em Nature of problem(approx. 50-250 words):}\\
Many problems in computational physics and applied mathematics are defined in vector spaces that can grow very rapidly with the problem size. This has been traditionally the case for quantum many-body systems, which require Hilbert spaces that grow exponentially large with the number of components of the physical model. However, this applies equally to many classical computations, such as the solution of large partial differential equations in high-dimensional spaces with very fine grids. In this context, developing efficient techniques that save \jj{computational resources, e.g., memory and time,} is a relevant challenge for areas such as condensed mattedr physics, quantum chemistry, and quantum information, as well as applications in fields such as machine learning, numerical analysis, and optimization.\\
{\em Solution method(approx. 50-250 words):}\\
The SeeMPS library provides a Python framework for quantum many-body and quantum-inspired methods based on Matrix Product States (MPS) and Tensor Trains (TT). These are tensor-network-based methods to quantize and compress the information associated elements in an exponentially large vector space. SeeMPS uses these techniques to develop efficient tensor and linear-algebra routines for diverse computational tasks such as eigenvalue searches and linear system solving, complemented by tools including Fourier transforms and other common numerical operations. The library prioritizes clarity, flexibility, and a modular architecture, making it suitable both for direct applications in state-of-the-art quantum physics problems, as well as for developing and testing quantum-inspired algorithms.
\end{small}
\end{abstract}
\end{frontmatter}
\clearpage

\begin{table}[H]
  \centering
  \bgroup
  \def\arraystretch{1.125}
  \begin{tabular}{|l|l|}
    \hline
    \textbf{Component} & \textbf{Implementation}\\ \hline\hline

    \multicolumn{2}{|c|}{\hyperref[sec:BLAS]{\textbf{``MPS-BLAS''}}}\\\hline
    \hyperref[sec:vector]{Vector} & $\mathbf{w},\mathbf{v} \in \mathrm{MPS}_\chi$\\\hline
    \hyperref[sec:matrix]{Matrix} & $U, Q\in\mathrm{MPO}_\chi$\\\hline
    \multirow{2}{*}{\hyperref[sec:simplification]{\jj{Compression}}} & $\mathrm{argmin}_{\mathbf{v}\in\mathrm{MPS}_\chi}\Vert\mathbf{v}-\mathbf{w}\Vert^2$ \\\cline{2-2} & Tensor cross-interpolation \\\hline
    \hyperref[sec:ewproduct]{Hadamard product} & $A^{s_i}_{\alpha\beta}B^{s_i}_{\gamma\delta}\to C^{s_i}_{(\alpha,\gamma),(\beta,\delta)}$ \\\hline
    \hyperref[sec:addition]{Vector addition} & $\mathrm{argmin}_{\mathbf{v}\in\mathrm{MPS}_\chi} \Vert\mathbf{v}- \sum_i w_i \mathbf{w}_i\Vert^2$\\\hline
    \hyperref[sec:matrix-vector]{Matrix-vector prod.} & $\mathrm{argmin}_{\mathbf{v}\in\mathrm{MPS}_\chi} \Vert\mathbf{v}- U \mathbf{w}\Vert^2$ \\\hline
    \hyperref[sec:outer]{Tensor product} & $\mathbf{v}\otimes\mathbf{w}$ \\ \hline
    \hline

    \multicolumn{2}{|c|}{\hyperref[sec:LAPACK]{\textbf{``MPS-LAPACK''}}}\\\hline
    \hyperref[sec:eigenvalue]{Eigenvalue search} & \hyperref[sec:power]{Power}, \hyperref[sec:Arnoldi]{Arnoldi}, \hyperref[sec:DMRG]{DMRG} \\\hline
    \hyperref[sec:linear-equation]{Linear system solver} & \hyperref[sec:CGS]{CGS, BiCGS}, \hyperref[sec:GMRES]{GMRES}, \hyperref[sec:DMRG-solve]{DMRG} \\\hline
    \hyperref[sec:QFT]{Fourier transform} & \hyperref[sec:QFT]{QFT MPO}\\\hline
    \hline

    \multicolumn{2}{|c|}{\hyperref[sec:function-representation]{\textbf{Functional Analysis}}}\\\hline
    \multirow{4}{*}{\hyperref[sec:loading]{Loading}} & \hyperref[sec:direct]{Direct constructions} \\\cline{2-2} & \hyperref[sec:orthogonal-polynomials]{Polynomial expansions} \\\cline{2-2}
                       & \hyperref[sec:TCI]{Tensor cross-interpolation (TCI)} \\\cline{2-2} & \hyperref[sec:complementary]{Complementary techniques} \\\hline
    \multirow{3}{*}{\hyperref[sec:differentiation]{Differentiation}} & \hyperref[sec:finite-differences]{Finite differences}\\\cline{2-2}
                       & \hyperref[sec:Fourier-differentiation]{Fourier differentiation}\\\cline{2-2}
                       & \hyperref[sec:HDAF]{HDAF}\\\hline
    \hyperref[sec:integration]{Integration} & \hyperref[sec:integration]{Newton--Cotes, Clenshaw--Curtis}\\\hline
    \multirow{2}{*}{\hyperref[sec:interpolation]{Interpolation}} & \hyperref[sec:fd_interpolation]{Finite differences} \\\cline{2-2}
                       & \hyperref[sec:fourier_interpolation]{Fourier interpolation} \\\hline
    \multirow{2}{*}{\hyperref[sec:PDE]{PDE Solution}} & Eigenvalue problems \\\cline{2-2}
                       & Source problems \\\hline
    \multirow{3}{*}{\hyperref[sec:evolution]{Evolution}} & \hyperref[sec:explicit]{Explicit Runge-Kutta} \\\cline{2-2}
                       & \hyperref[sec:implicit]{Implicit Crank-Nicolson, Radau} \\\cline{2-2}
                       & \hyperref[sec:TDVP]{TDVP} \\\hline
    \hline

    \multicolumn{2}{|c|}{\hyperref[sec:quantum]{\textbf{Quantum Many-Body Physics \& Computing}}}\\\hline
    \multirow{2}{*}{\hyperref[sec:Hamiltonian]{Hamiltonian MPO}} & \rtt{api/class/seemps.hamiltonians.InteractionGraph}{InteractionGraph} \\\cline{2-2}
                       & \hyperref[sec:TEBD]{Nearest-neighbor problems} \\\hline
    \hyperref[sec:eigenvalue]{Ground state search} & \hyperref[sec:DMRG]{DMRG} \\\hline
    \multirow{2}{*}{\hyperref[sec:QFT]{Time evolution}} & \hyperref[sec:evolution]{Same as above}\\\cline{2-2}
   & \hyperref[sec:TEBD]{TEBD}\\\hline
    \hyperref[sec:quantum-computer]{Quantum circuits}
    & Parameterized quantum circuits\\\hline
  \end{tabular}
  \egroup
  \caption{Summary of fundamental operations, intermediate algorithms, and high-level applications contained in SeeMPS.}
  \label{tab:algorithms}
\end{table}

\newpage
\section{Introduction}
Many problems in computational physics, applied mathematics, and scientific computing are naturally formulated in vector spaces whose dimensionality grows exponentially with system size or resolution. This situation is ubiquitous in quantum many-body physics, where Hilbert spaces grow exponentially with the number of degrees of freedom. However, the same challenge arises in classical problems, such as the numerical solution of high-dimensional partial differential equations or multivariate interpolation over very fine grids. In all these cases, direct representations rapidly become impractical, motivating the use of compressed representations that exploit structure, locality, and correlations.

Tensor network methods provide a systematic framework to address the curse of dimensionality. Among them, Matrix Product States (MPS)—known in numerical analysis as Tensor Trains (TT)—and their operator counterpart, Matrix Product Operators (MPO), stand out due to their favorable balance between expressive power and computational efficiency. Originally found in the context of the Density Matrix Renormalization Group (DMRG) for one-dimensional quantum lattice systems~\cite{white1992, white1993, schollwock2011}, MPS representations evolved into a well-understood framework for algorithm development, including solving large-scale eigenvalue problems~\cite{verstraete2004a, porras2006} and simulating time-evolving quantum many-body systems~\cite{vidal2004, verstraete2004, garcia-ripoll2006, haegeman2011}. However, in parallel to this, MPS entered other domains as efficient compressed representations of large datasets. Thus, we find early works studying quantum and quantum-inspired image processing~\cite{latorre2005}, encoding of multidimensional functions~\cite{iblisdir2007}, and the solution of large-scale PDEs~\cite{lubasch2018, garcia-ripoll2021, gourianov2022}. These efforts have parallel developments in the domain of applied mathematics, where MPS have been discovered under the name of tensor trains~\cite{oseledets2011} (TT) or quantics (quantized) tensor trains~\cite{oseledets2010, khoromskij2011, khoromskij2011a} (QTT),  finding applications in the domain of high-dimensional function representation, integration, interpolation, etc.

Despite their widespread use, the practical adoption of MPS/TT methods across disciplines remains hindered by fragmented software ecosystems. Existing libraries often focus either on highly specialized quantum many-body simulations or on low-level tensor operations, leaving users to re-implement basic linear algebra routines, truncation strategies, and error control mechanisms. As a result, MPS/TT objects are rarely treated as first-class numerical entities comparable to vectors and matrices in conventional linear algebra libraries. The SeeMPS library addresses this gap by providing a unified Python framework in which MPS and MPOs are treated as elements of a finite-precision linear algebra~\cite{garcia-molina2024}, offering the user an expressive language to implement a wide variety of efficient algorithms.

In this work, we introduce the SeeMPS library, a Python library implementing the MPS/TT formalism. The design of the library is guided by three principles. First, tensor networks are used systematically as representations of vectors and linear operators acting on exponentially large vector spaces. Second, a well-defined set of fundamental operations—an “MPS-BLAS” in analogy with the BLAS standard—is identified and implemented, including scaling, addition, inner products, matrix–vector products, and element-wise operations, all equipped with controlled truncation and error estimates. Third, these low-level primitives are composed into higher-level algorithms—an MPS analogue of LAPACK—supporting eigenvalue searches, linear system solvers, and Fourier transforms within a unified abstraction.

Beyond linear algebra, SeeMPS extends the MPS/TT formalism to function representation and numerical analysis. Multidimensional functions discretized on exponentially large grids are encoded using quantized tensor trains (QTT), following ideas found in the quantum physics and quantum computing world~\cite{zalka1998, grover2002, lubasch2018, garcia-ripoll2021}, as well as in the field of numerical analysis~\cite{oseledets2010, khoromskij2010, khoromskij2011}. The library provides multiple strategies to construct these representations, including direct analytic constructions, polynomial expansions~\cite{holzner2011, halimeh2015, lindsey2024, rodriguez-aldavero2025}, Fourier-based methods~\cite{garcia-ripoll2021, chen2023, garcia-molina2022}, and sampling-based tensor cross-interpolation algorithms~\cite{oseledets2010a, savostyanov2011, mikhalev2018, dolgov2020, nunez-fernandez2022, nunez-fernandez2025, sozykin2022, ritter2024}. On top of these encodings, SeeMPS implements numerical differentiation, interpolation, integration, and time evolution using MPO-based operators, enabling efficient solvers for ordinary and partial differential equations in high dimensions.

The scope of SeeMPS's applications spans quantum many-body physics, quantum-inspired numerical analysis, and high-dimensional scientific computing. While the library includes established tools such as DMRG, time-evolution algorithms, and quantum circuit emulation, its architecture is intentionally general: MPS are treated as compressed vectors and MPOs as compressed linear operators, independently of any specific physical interpretation. This perspective facilitates the transfer of methods between different application domains and lowers the barrier for adopting tensor-network techniques outside their original quantum-physics context.

The organization of this article is motivated by the library structure described in Table~\ref{tab:algorithms}. Section~\ref{sec:BLAS} introduces the \jj{basic linear algebra package} based on MPS and MPO representations for vectors and matrices and finite-precision operations, which we call MPS-BLAS. Section~\ref{sec:LAPACK} describes how these basic operations are combined into higher-level linear algebra algorithms (MPS-LAPACK), including eigenvalue solvers, linear system solvers, and Fourier transforms. Sections~\ref{sec:function-representation} and \ref{sec:differentiation} focus on function encoding, interpolation, differentiation, and integration within the MPS/TT framework. Subsequent sections illustrate applications ranging from the numerical solution of differential equations (Sect.~\ref{sec:PDE}) to applications in quantum many-body physics and quantum computing (Sect.~\ref{sec:quantum}). We close this work with a discussion of the library's architecture in Sect.~\ref{sec:architecture} and some general conclusions and outlook in Sect.~\ref{sec:conclusion}.

\section{Basic Linear Algebra Operations}
\label{sec:BLAS}
The SeeMPS library is based on an approach towards using tensor networks with three guiding principles. The core principle is the use of tensor networks as representations of points and of linear transformations in a vector space, to compress the information and speed up linear (and nonlinear) algebra operations. The second principle is the identification of a battery of fundamental operations---what we call an MPS-BLAS in analogy with the traditional BLAS library and standard---that can support all other high-level algorithms. These two principles are combined with a design choice, which is the selection of the so-called \textit{matrix product state (MPS)} or \textit{tensor train (TT)} as the core structure with which to represent the elements of the vector space and to implement the fundamental operations.

\subsection{Vector Quantization: MPS / TT}
\label{sec:vector}
\begin{figure}[t]
  \centering
  \includegraphics[width=0.7\linewidth]{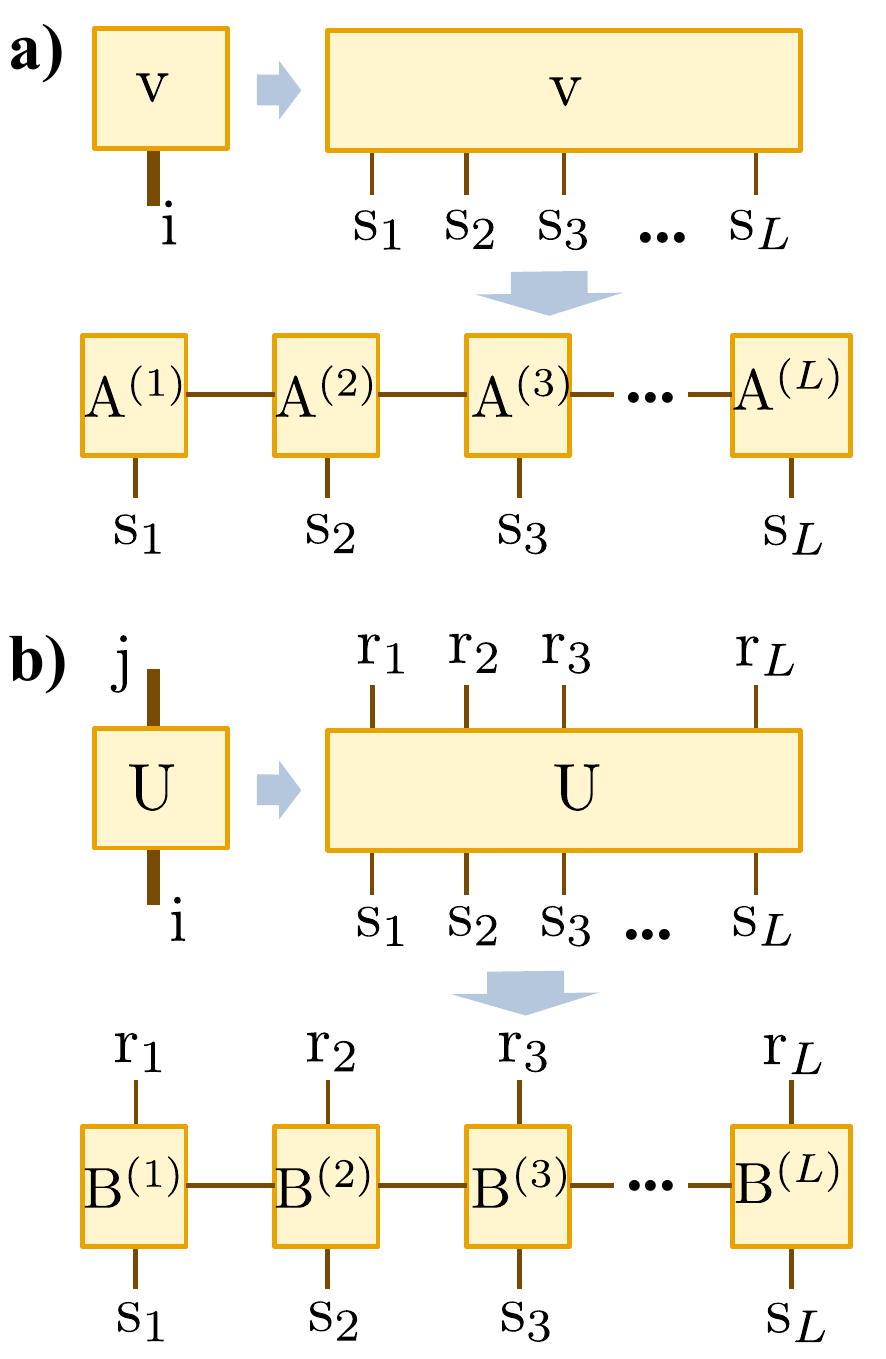}
  \caption{(a) A high-dimensional vector $v_i$ is reinterpreted as a tensor $v_{s_1,s_2,\ldots,s_{\jj{L}}}$ that admits a matrix product state (MPS) or tensor-train (TT) representation with smaller tensors $A_{\alpha_{n-1},\alpha_{n}}^{(n),\jj{s_n}}$\jj{. For broad classes of vectors the bond dimensions remain small, so that the representation occupies less memory while approximating the vector to within a controlled tolerance, in many cases close to machine precision~\cite{garcia-ripoll2021, jobst2024}}. (b) A similar procedure applied to matrices leads to what is known as a matrix product operator (MPO).}
  \label{fig:tensors}
\end{figure}
Many problems in physics and applied mathematics are formulated using vector spaces. Typically, a datapoint or vector is represented in an orthonormal basis $\mathbf{v}=\sum_iv_i\mathbf{e}_i$ as an array mapping integer values to real or complex values, e.g.
\begin{equation}
  \label{eq:vector}
  v_i : i \in {1, 2,\ldots, D} \to \mathbb{C}.
\end{equation}
Quantum Mechanics is a relevant example of these applications, because \textit{all physical states and all physical operations} are uniquely adscribed to vectors and linear transformations in a Hilbert space of varying dimensionality.

In many fields, as in Quantum Mechanics, one often finds that the vector space admits a tensor product structure, that is, the \textit{system it represents is decomposable} into smaller units. This applies to models of electrons hopping in a lattice, \jj{atoms in an ion trap,} or qubits in a quantum computer. In that case, the vector space splits into $L$ different components, often with the same size, $\mathbf{v}\in \mathcal{H}=\mathcal{H}_1\otimes \jj{\cdots\otimes} \mathcal{H}_L$.

In this scenario, it makes sense to choose an orthonormal basis that is the tensor product of the basis used to represent the individual components,
\begin{align}
  \mathbf{e}_i &:= \mathbf{e}_{s_1}\otimes \cdots \jj{\otimes} \mathbf{e}_{s_L},\\
  \mathbf{v} &= \sum_{s_{1},\ldots,s_L}v_{s_1,s_2,\ldots, s_L}\mathbf{e}_{s_1}\otimes \cdots \jj{\otimes} \mathbf{e}_{s_L}.
\end{align}
In doing so, we also have established two maps, one between integer labels in a finite domain $i\in{1,\ldots,d^L}$ to a vector of integers that run over smaller domains, $s_i\in\{1,2,\ldots,d_i\}$, and its inverse. Assuming \textit{most significant indices} come first
\begin{align}
  i & \to \mathbf{s} = (s_{1},s_{2},\ldots, s_{L}),\label{eq:quantization}\\
  \mbox{with}~ i&:= \sum_{\jj{k}} s_{\jj{k}} \prod_{j>\jj{k}}d_j,\notag
\end{align}
we have a recipe to tensorize a vector~\eqref{eq:tensorization}
\begin{align}
  v_i & \to v_{s_{1},s_{2},\ldots,s_{L}}.\label{eq:tensorization}
\end{align}
A particular instance of this representation is that where we decompose a vector of dimension $2^n$ into $n$ subspaces with two-dimensional objects or ``qubits''. We refer to this case as a \textit{quantization} of the coordinate and the resulting tensor train as a \textit{quantized tensor train} (see Sect.~\ref{sec:function-representation}). \jj{Throughout this work we reserve $L$ for the number of MPS tensors and $d$ for the local dimension of each index, so that a generic vector has size $d^L$; in the quantized binary case $d=2$ and we write $n$ for the number of qubits (hence $n=L$ tensors).}

This equivalence between a vector and a tensor is useful because it opens the door to a variety of compressed representations that exploit hidden correlations in the resulting tensors. Known as \textit{tensor networks}, these representations split the map $v_{s_1,s_2,\ldots,s_L}$ as a graph of smaller tensors that are contracted with each other, to produce the desired values. In SeeMPS, we rely on a particular tensor network that receives the names \textit{matrix product state} (MPS) or \textit{tensor train} (TT) in the fields of quantum physics and applied mathematics. This representation reconstructs each value $v_{s_1,s_2,\ldots,s_L}$ as a trace over a product of $L$ matrices, each labeled by a separate index
\begin{align}
  \label{eq:general-MPS}
  v_{s_1,s_2,\ldots,s_L}
  &= \sum_{\alpha_1,\ldots, \alpha_L}A^{(1)}_{\alpha_Ls_1\alpha_1} A^{(2)}_{\alpha_1s_2\alpha_2}\cdots A^{(L)}_{\alpha_{L-1}s_L\alpha_L} \\
  &=\mathrm{tr}\left(\mathbf{A}^{(1)s_1}\mathbf{A}^{(2)s_2}\cdots \mathbf{A}^{(L)s_L}\right).\notag
\end{align}
This change from a vector $\textbf{v}$, to a tensor $v_{s_1,s_2,\ldots,s_L}$ and to a tensor train is sketched in Fig.~\ref{fig:tensors}(a).

In many applications, the tensors $\mathbf{A}^{s_i}$ are small, and the intermediate indices $1 \leq \alpha_i < \chi$ lie below some small limit $\chi$ known as the \textit{bond dimension}. In that case, the MPS/TT actually compresses the vector, which goes down from a size $d^L$ to \jj{at most $Ld\chi^2$ (an upper bound, saturated only with periodic boundary conditions)}. This memory reduction also translates into a compression in time when we implement other algorithms, as discussed below.

Finally, some practical matters. First, in SeeMPS we adopt a common convention that the first and last indices are fixed to $\alpha_L=1$ (and dropped from the picture, as in Fig.~\ref{fig:tensors}), to simplify algorithms and improve efficiency. \jj{Fixing these open boundaries to dimension one removes the periodic-trace bookkeeping and guarantees a well-defined isometry center, on which the canonical form and all sweep-based algorithms below rely.}  Second, now our original vector $v_i$ becomes a collection of tensors, which in SeeMPS we store as a specialized \rtt{seemps_objects_mps}{MPS} Python sequence object that only contains NumPy arrays with 3 indices
\begin{equation}
  A_{\alpha_{n-1}s_n\alpha_n}^{(n)} \to \mathtt{A[n][\alpha_{n-1},s_n,\alpha_n]}.
\end{equation}

\subsection{Matrix Representation: MPO}
\label{sec:matrix}
Just like a vector admits a tensor decomposition, linear operators or matrices can also be compressed in a similar form. Figure~\ref{fig:tensors}(b) sketches how a matrix with two indices, $U_{ji}$, when acting on a composite vector space, becomes a tensor with $2L$ indices, that can be split into $L$ smaller tensors, a set of $L$ matrices labeled by pairs of subindices
\begin{align}
  \label{eq:MPO}
  U_{ji} &\to U^{j_1j_2\cdots j_L}_{i_1i_2\ldots i_L}
  = \sum_{\beta_1,\ldots,\beta_L}B^{(1)}_{\beta_Lj_1i_1\beta_1}\cdots B^{(L)}_{\beta_{L-1}j_Li_L\beta_L}\notag\\
 &= \mathrm{tr}\left(\mathbf{B}^{(1),j_1i_1}\mathbf{B}^{(2),j_2i_2}\cdots \mathbf{B}^{(L),j_Li_L}\right).
\end{align}
This decomposition is known in quantum physics as a matrix product operator (MPO)\jj{---and, equivalently, as a \textit{tensor-train operator} (TTO) in the numerical-analysis lit\-erature---} and is implemented in SeeMPS as a \rtt{seemps_objects_mpo}{MPO} sequence object
\begin{equation}
  B^{(n)}_{\beta_{n-1}j_ni_n\beta_n} \to \mathtt{B[n][\beta_{n-1},j_n,i_n,\beta_n]}.
\end{equation}

\subsection{\jj{Compression} Techniques and Error Estimates}
\label{sec:simplification}
Most algorithms in SeeMPS involve controlled transformations of the tensors in the MPS or MPO objects. This is the case when we add up two MPS (Sect.~\ref{sec:addition}), when we contract MPO and MPS (Sect. \ref{sec:matrix-vector}), or when we apply higher-level algorithms for evolution (Sect.~\ref{sec:evolution}) or \jj{an} eigenvalue search (Sect.~\ref{sec:eigenvalue}). These transformations may involve a rather uncontrolled growth of the bond dimensions $\chi$---an increase that may become exponentially larger in most cases. Fortunately, in many useful situations, one can find new choices of tensors that lead to more compressed representations. SeeMPS offers two compression algorithms.

The first \jj{\textit{compression algorithm}} creates an MPS in \textit{canonical form}~\cite{cirac2021}. This is a \rtt{seemps_objects_canonical}{CanonicalMPS} object with an explicit index $0\leq k<L$, such that tensors $A^{(n<k)}$ and $A^{(k<n)}$ are respectively left- and right-isometries. \jj{A tensor is a \textit{left isometry} when $\sum_{\alpha_{n-1},s_n}A^{*}_{\alpha_{n-1}s_n\alpha_n}A_{\alpha_{n-1}s_n\alpha_n'}=\delta_{\alpha_n\alpha_n'}$ (and analogously for a right isometry, contracting the physical and right-bond indices); geometrically, these tensors map an orthonormal basis of one bond space into an orthonormal set on the other, so that partial contractions of the network preserve norms and all inner products and expectation values can be evaluated locally, from the center tensor $A^{(k)}$ alone.} The algorithm to create these isometries is based on an iterative singular value decomposition of the tensors that only keeps singular values above a given tolerance. Imposing a \textit{maximum bond dimension} $\chi_\text{max}$ and a \textit{truncation tolerance} $\varepsilon$, is equivalent to keeping at most $\chi_\text{max}$ singular values whose magnitude lies above $\varepsilon_\text{tol}$ times the largest singular value. This procedure limits the size of the resulting isometries and achieves a natural compression of the state with a measurable \textit{truncation error}, determined by the magnitude of the dropped singular values.

The second and most powerful \jj{compression} algorithm (used by default everywhere) is a variational search for optimal tensors that minimize the distance to the MPS vector to be compressed $\phi$
\begin{equation}
  \label{eq:simplification}
  \psi_\text{opt} =  \mathrm{argmin}_{\psi \in \mathrm{MPS}_\chi}
  \left\Vert\psi - \phi\right\Vert^2.
\end{equation}
This search happens iteratively through a sequential update of pairs of tensors in the MPS structure~\cite{garcia-ripoll2006}, starting from a good guess and producing another \texttt{CanonicalMPS}. Once more, the optimization may be driven by a \textit{maximum bond dimension} \jj{(the truncation parameter $\chi_\text{max}$ introduced earlier in this section)} and an \textit{error tolerance}, or by both. In all cases, the algorithm finds a candidate with an estimate of the total \textit{truncation error}. Note also that both \jj{compression} algorithms exist for MPS and MPO in different functions,  \rtt{algorithms/mps_simplification}{simplify} and \rtt{algorithms/mps_simplification}{simplify\_mpo}. \jj{Throughout this work we refer to this operation as \textit{compression}, following the standard terminology of the tensor-network community; for backward compatibility, the corresponding SeeMPS routines retain their historical names \texttt{simplify} and \texttt{simplify\_mpo}.}

SeeMPS activates \jj{compression} steps implicitly in most matrix-vector operations (Sect.~\ref{sec:matrix-vector}) and algorithms that take an optional \texttt{strategy} argument. An upper bound for the accumulated truncation error of all operations that act on a vector is kept in the field \texttt{error} of the MPS object, which provides an estimate of the worst case scenario for this imprecise algebra. For instance, the \rtt{api/function/seemps.evolution.runge_kutta}{runge\_kutta} routine takes a \texttt{strategy} option that determines which \jj{compression} routine to use after each evolution step. This argument is a \texttt{Strategy} object that determines the \jj{compression} algorithm, maximum bond dimension, and tolerances. As the algorithm changes the underlying vector, the accumulated errors are stored in the MPS as a crude witness of possible deterioration of the approximation. Though in practice the results may be better than the estimates~\cite{garcia-ripoll2006}.

In addition to these two implicit algorithms, SeeMPS offers a third family of \jj{compression} mechanisms based on tensor cross interpolation (TCI). These efficient but potentially less accurate methods must be manually invoked using a different functional API (c.f. Sect.~\ref{sec:TCI}).

\subsection{Basic Linear Algebra Subsystem (BLAS)}
\begin{figure}[t]
  \centering
  \includegraphics[width=0.75\linewidth]{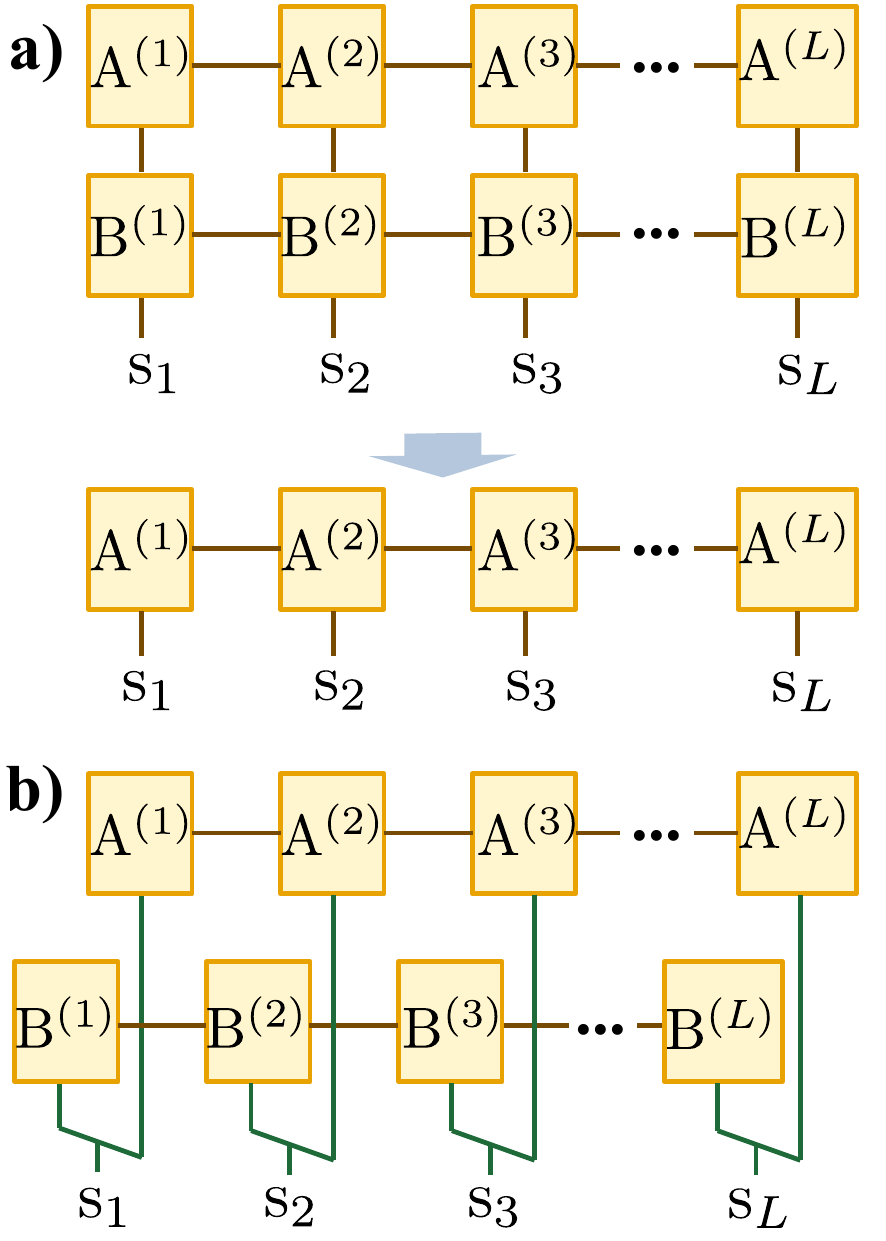}
  \caption{\label{fig:operations}Illustration of core MPS-BLAS operations in SeeMPS: (a) matrix-vector product between an MPS with tensors $\{A^{(n)}\}_{n=1}^L$ and an MPO with tensors $\{B^{(n)}\}_{n=1}^L$; (b) element-wise vector products between two MPS with tensors $\{A^{(n)}\}_{n=1}^L$ and $\{B^{(n)}\}_{n=1}^L$.}
\end{figure}
The MPS and MPO classes are endowed with a complete basic linear algebra package of operations acting on 1D tensor networks. More precisely, what we implement is a finite-precision BLAS (as described first in Ref.~\cite{garcia-molina2024}), where the operations are controlled in error and tensor size with the algorithms discussed in Sect.~\ref{sec:simplification}. \jj{Here \textit{finite precision} denotes an algebra in which every operation carries a controlled truncation error---in analogy with the rounding of floating-point BLAS---as opposed to the exact arithmetic of symbolic computation. It is not a statement about the floating-point format, which is double precision throughout (Sect.~\ref{sec:architecture}).}

\subsubsection{Scaling and Addition}
\label{sec:addition}
Both \texttt{MPS} and \texttt{MPO} objects can be multiplied or divided by a scalar. This is achieved by overloading the multiplication and division operators \texttt{*} and \texttt{/}, through the Python methods \texttt{\_\_mul\_\_} and \texttt{\_\_truediv\_\_}. We have found that numerical stability is improved by spreading the factors over all tensors.

The Python addition \texttt{+} and subtraction \texttt{-} operators also act on  \texttt{MPS} and \texttt{MPO} objects. These operators are lazy: instead of creating new vectors and operators, linear combinations of vectors and operators are stored in \rtt{seemps_objects_sum}{MPSSum} and \texttt{MPOSum} objects, respectively. This avoids creating large tensors and enables doing multiple sums. Thus, a linear combination \texttt{a*A+ b*B+ c*C + $\ldots$} will be stored as a single object that preserves the weights \texttt{[a, b, c, $\ldots$]} and the summands \texttt{[A, B, C, $\ldots$]}.

To make these lazy operations useful, the library strives to accept \texttt{MPSSum} and \texttt{MPOSum} objects wherever a vector or an operator is required. However, in many situations, these objects must be converted back into MPS and MPO objects to avoid a combinatorial explosion of intermediate terms. This transformation can be eagerly implemented using the \texttt{join()} method, which, given a lazy sum, creates a tensor product structure of the tensors in the sums. Since this may create very large objects, most SeeMPS algorithms use instead the compression routines  \rtt{algorithms/mps_simplification}{simplify} and \rtt{algorithms/mps_simplification}{simplify\_mpo}, which \jj{compress} the sum either by solving the variational problem
\begin{equation}
\label{eq:mps-sum-opt}
\psi_\mathrm{opt} =  \mathrm{argmin}_{\psi \in \mathrm{MPS}_\chi}
\left\Vert \psi - \sum_{l=1}^L \phi_l \right\Vert^2,
\end{equation}
or using a \jj{two-site} canonicalization step (c.f. Sect.~\ref{sec:simplification}).

\subsubsection{Matrix-vector Product}
\label{sec:matrix-vector}
The action of an MPO $U$ (with tensors $\{\mathbf{B}^{(n)}\}$) onto an MPS $\textbf{v}$ (with tensors $\{\mathbf{A}^{(n)}\}$) can be easily written as an MPS with enlarged tensors $\mathbf{C}^{(n)}$ (c.f. Figure~\ref{fig:operations}a)
\begin{equation}
  \label{eq:MPO-times-MPS}
  C_{\gamma_{n-1}j_n\gamma_n}=\sum_{i_n}
B_{\beta_{n-1}\beta_n}^{j_ni_n}A_{\alpha_{n-1}\alpha_n}^{i_n}.
\end{equation}
The aggregated tensor combines pairs of indices $\gamma_n=(\beta_n,\alpha_n)$ as a single integer with enlarged range $\gamma_n = \beta_n d_n + \alpha_n$, where $d_n = \max \alpha_n$ is the dimension of the fastest running index. This notation is used elsewhere in this work to denote other tensor compositions.

In SeeMPS, the matrix-vector multiplication is implemented by overloading NumPy's operator \texttt{@}\jj{, with both orders available: \texttt{U @ v} applies the operator to the state from the left, while the reverse order \texttt{v @ U} applies it from the right, as is often convenient in quantum applications}. However, instead of just contracting the tensors, the library allows the MPO object to have an implicit \jj{\textit{compression strategy}}. According to this, when the variational algorithm is selected, the actual multiplication amounts to solving the problem
\begin{equation}
  \mathrm{argmin}_{\mathbf{w}\in\mathrm{MPS}_\chi}\left\Vert\mathbf{w} - U \mathbf{v}\right\Vert^2,
\end{equation}
with the prescribed truncation and \jj{compression} tolerances, and possible bond dimension limits.

\subsubsection{Matrix-matrix Product}
\label{sec:matrix-matrix}
SeeMPS also overloads NumPy's tensor product operator \texttt{@} to support the multiplication of MPOs. This is a lazy operation that results in an \rtt{api/class/seemps.operators.MPOList}{MPOList} object that stores a list of MPO factors. \jj{Since this object represents a lazy \emph{product} of operators, it is also exposed under the alias \texttt{MPOProd}, which we intend to adopt as the canonical name in a future release.} The \texttt{MPOList} can be used in two equivalent ways. First, the individual MPOs can be combined into a single MPO using the \texttt{join} method, yielding an operator that represents the product of all MPOs in the list, but whose size may grow exponentially.

Second, the MPOs can be applied sequentially to an MPS, producing intermediate MPS representations after each application. This option is used almost everywhere in SeeMPS because it allows a controlled growth of bond dimensions thanks to the intermediate \jj{compressions} that are activated after each MPO-MPS product.

A relevant example of \texttt{MPOList} is the Quantum Fourier Transform, an efficient routine that is implemented using a product of unitary layers that implement the QFT quantum circuit (c.f. Sect.~\ref{sec:QFT}). This object is always applied sequentially on the vector to be transformed, to ensure a smaller size of the intermediate products.

\subsubsection{Inner Product}
\label{sec:inner-product}
The inner product or scalar product of two vectors in MPS representation, denoted $\langle\mathbf{v},\mathbf{u}\rangle$ or $\mathbf{v}^\dagger\mathbf{v}$, is implemented by the function \rtt{api/function/seemps.state.scprod}{scprod} and its NumPy compatible alias \rtt{api/function/seemps.state.vdot}{vdot}.
The scalar product of two MPS is a very efficient operation that can be implemented as a sequence of matrix-vector multiplications with a cost that scales as $\jj{\mathcal{O}(L\chi^3 d)}$. \jj{In particular, the squared norm $\langle\mathbf{v},\mathbf{v}\rangle$ of an MPS in canonical form is obtained from its center tensor alone, at a size-independent cost $\mathcal{O}(d\chi^2)$.}

For quantum applications, the library helps \jj{to} compute the expected value of operators acting on subsystems of the vector space. For instance, \rtt{api/function/seemps.expectation.expectation1}{expectation1(v, O, i)} computes the expected value $\langle\mathbf{v}, O_i\mathbf{v}\rangle$ with an observable $O_i=\mathbb{I}^{\otimes(i-1)}\otimes O \otimes \mathbb{I}^{\otimes(L-i)}$ that acts on the $i$-th subsystem of the Hilbert space. Other functions efficiently compute bipartite correlations over pairs of subsystems, or over all possible subsystems in the space, as well as an \rtt{api/class/seemps.operators.MPO}{MPO.expectation} method to estimate the projection of an operator onto one or two vectors\jj{,} e.g., \texttt{Q.expectation(u,v)} produces $\langle\mathbf{u}, Q \mathbf{v}\rangle$.

\subsubsection{Element-wise Products}
\label{sec:ewproduct}
Motivated by applications in the study of nonlinear equations, SeeMPS also offers the element-wise product or Hadamard product of two vectors
$\mathbf{v}, \mathbf{w} \in \mathbb{C}^n$
\begin{equation}
\mathbf{v} \odot \mathbf{w}
= (v_1 w_1, v_2 w_2, \dots, v_n w_n).
\end{equation}
\jj{Writing each entry of the two operands as a product of matrices, $v_{s_1\cdots s_L}=A^{s_1}\cdots A^{s_L}$ and $w_{s_1\cdots s_L}=B^{s_1}\cdots B^{s_L}$, the corresponding entry of the Hadamard product factorizes site by site,}
\begin{align}
  \jj{(\mathbf{v}\odot\mathbf{w})_{s_1\cdots s_L}}
  &\jj{= \left(A^{s_1}\cdots A^{s_L}\right)\left(B^{s_1}\cdots B^{s_L}\right)}\\
  \notag&\jj{= \left(A^{s_1}\otimes B^{s_1}\right)\cdots\left(A^{s_L}\otimes B^{s_L}\right),}
\end{align}
\jj{where the last equality follows from the mixed-product property of the Kronecker product $\otimes$. The element-wise product is therefore again an MPS, whose tensors are the local Kronecker products $C^{s_n}=A^{s_n}\otimes B^{s_n}$; that is, joining $A_{\alpha_{n-1}s_n\alpha_n}$ and $B_{\beta_{n-1}s_n\beta_{n}}$ into}
\begin{align}
  C_{\gamma_{n-1}s_n\gamma_{n}}
  &= C_{(\alpha_{n-1},\beta_{n-1})s_n(\alpha_n,\beta_n)}\\
  &=A_{\alpha_{n-1}s_n\alpha_n}B_{\beta_{n-1}s_n\beta_{n}},\notag
\end{align}
as sketched in Figure~\ref{fig:operations}b. \jj{The two bond indices are merged into the single combined index $\gamma_n=(\alpha_n,\beta_n)$, packed into a single integer $\gamma_n=\alpha_n\chi_B+\beta_n$ following the convention introduced for the matrix--vector product above. The resulting MPS therefore carries a bond dimension equal to the product $\chi_A\chi_B$ of the two operands (before any subsequent compression)}. In the library, this element-wise product is implemented by overloading the Python operator $\texttt{*}$ to act among \texttt{MPS} and \text{MPSSum}.

\subsubsection{Outer Operations}
\label{sec:outer}
\begin{figure}[t]
  \centering
  \includegraphics[width=0.8\linewidth]{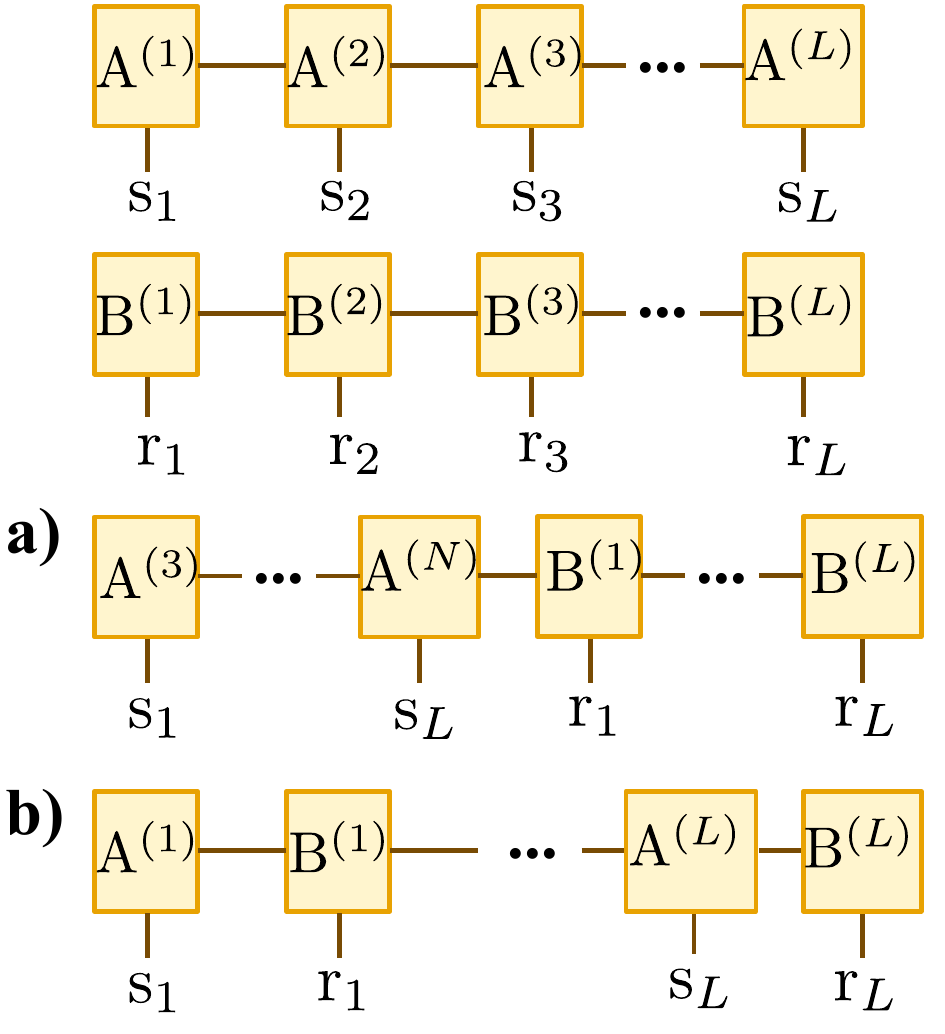}
  \caption{The tensor product of two vectors $\textbf{v}\otimes\textbf{w}$ represented by tensor trains $\{\mathbf{A}\}$ and $\{\mathbf{B}\}$ is constructed in SeeMPS using (a) a sequential \texttt{``A''} or an (b) interleaved \texttt{``B''} order.}
 \label{fig:tensorization}
\end{figure}
We include here \rtt{api/function/seemps.state.mps_tensor_product}{mps\_tensor\_product} and its sibling \rtt{api/function/seemps.state.mps_tensor_sum}{mps\_tensor\_sum}, which correspond to the tensorization $\mathbf{v}\otimes\mathbf{w}$ and the sum of extended vectors $\mathbf{v}\otimes\mathbf{e}_\mathbf{w} +\mathbf{e}_\mathbf{v}\otimes\mathbf{w}$, where $\mathbf{e}_\mathbf{x}$ is a vector of ones with the same size as the argument $\mathbf{x}$.

\begin{figure*}[t]
  \centering
  \includegraphics[width=0.85\linewidth]{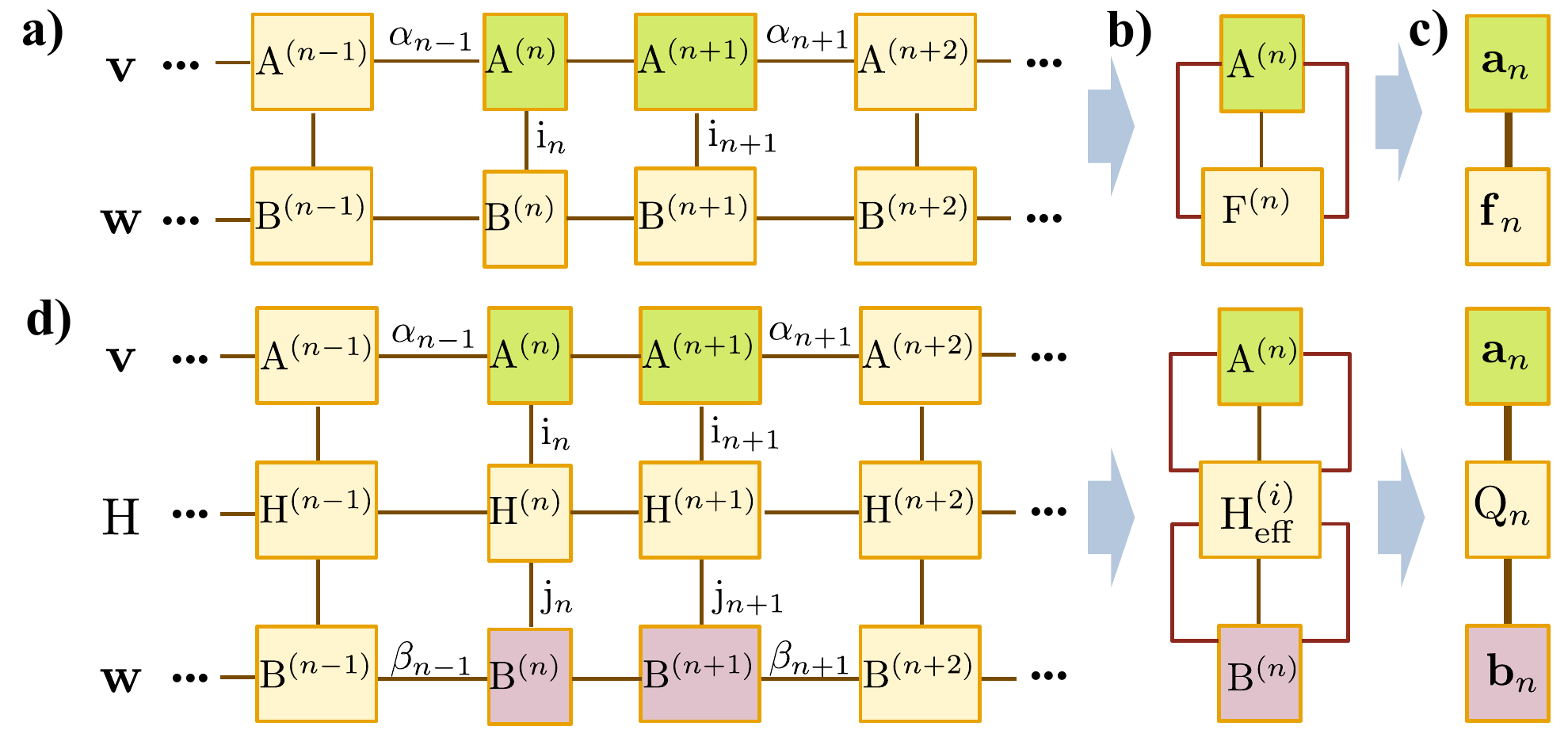}
  \caption{\label{fig:forms}\jj{Tensor-network two-site reduction of the linear~\eqref{eq:single-site-linear} and quadratic forms~\eqref{eq:single-site-quadratic} of Sect.~\ref{sec:forms} to standard DMRG environments. (a) Linear form. The inner product $\langle\mathbf{v},\mathbf{w}\rangle$ of two MPS with tensors $A^{(i)}$ and $B^{(i)}$ is the contraction of the two-layer network in~(a). Singling out the active tensor $A^{(i)}$ and contracting all the others collapses the rest of the network into an effective tensor $B^{(i)}_{\mathrm{eff}}$---the environment, drawn in red---in~(b), so that the inner product becomes a scalar product $\mathbf{a}^{(i)\dagger}\mathbf{b}^{(i)}$ between the two effective vectors of~(c) [cf.\ Eq.~\eqref{eq:single-site-linear}]. (d) Quadratic form. For the expectation value $\langle\mathbf{v},H\mathbf{v}\rangle$ of an MPO $H$, the same reduction collapses the surrounding tensors into an effective matrix $Q^{(i)}$, turning the expectation value into a quadratic form $\mathbf{a}^{(i)\dagger}Q^{(i)}\mathbf{a}^{(i)}$ [cf.\ Eq.~\eqref{eq:single-site-quadratic}].}}
\end{figure*}
Conceptually, both operations are performed by embedding the input MPS into a common vector space of a higher dimension, followed by possible additions. If we label $\{\mathbf{e}_{s}\}$ and $\{\mathbf{e}_{r}\}$ the basis on which two vectors $\mathbf{v}$ and $\mathbf{w}$ are defined, in SeeMPS, their tensor product may happen with two different orders:
\begin{align}
  \mbox{Order \texttt{"A"}:}~
  & \mathbf{e}_{s_1}\otimes\cdots\mathbf{e}_{s_{\jj{L}}} \otimes
    \mathbf{e}_{r_1}\otimes\cdots\mathbf{e}_{r_{\jj{L}}}\\
  \mbox{Order \texttt{"B"}:}~
  & \mathbf{e}_{s_1}\otimes\mathbf{e}_{r_1}\otimes\cdots\otimes\mathbf{e}_{s_{\jj{L}}} \otimes
    \mathbf{e}_{r_{\jj{L}}}.
\end{align}
These two orders result in a reorganization of the tensors in the final MPS (c.f. Fig.~\ref{fig:tensorization}). \jj{Order \texttt{"A"} is an exact reindexing of the tensor product at the existing bond dimension, while order \texttt{"B"} groups together digits of the same scale across coordinates. Neither order dominates universally: for smooth, correlated multivariate functions the interleaved order \texttt{"B"} typically yields a lower bond dimension after compression~\cite{garcia-ripoll2021, rodriguez-aldavero2025}, whereas for separable or weakly correlated functions it can be larger than the sequential order \texttt{"A"}. The most efficient choice is therefore function-dependent.}

\subsection{Linear and Quadratic Forms}
\label{sec:forms}
Many MPS algorithms---some mentioned before and others described below---depend on the estimation and optimization of functions that are linear and quadratic in their input vectors. SeeMPS recognizes this and offers two abstractions that ease and optimize our manipulation of those formulas. Given two MPS vectors $\mathbf{v}$ and $\mathbf{w}$ with tensors $\mathbf{A}^{(n)}$ and $\mathbf{B}^{(n)}$, and an MPO $H$, the \texttt{AntilinearForm} and the \texttt{QuadraticForm} objects allow us to focus on the $(n,n+1)$ pair of tensors and write
\begin{align}
  \label{eq:single-site-linear}
  \langle{\mathbf{v},\mathbf{w}}\rangle
  &=\mathbf{a}_n^\dagger \mathbf{f}_n
    \\
  \label{eq:single-site-quadratic}
  \langle{\mathbf{v},H\mathbf{w}}\rangle
  &= \mathbf{a}^\dagger_n Q_n \mathbf{b}_n.\notag
\end{align}
with an antilinear $\mathbf{f}$ and a quadratic form $Q$ that are efficiently computable, and two artificial vectors that result from joining pairs of neighboring tensors
\begin{align}
  (\mathbf{a}_n)_{(\alpha_{n-1},i_n,i_{n+1},\alpha_{n+1})}
   & := \sum_{\alpha_n}A_{\alpha_{n-1}i_n\alpha_n}A_{\alpha_{n}i_{n+1}\alpha_{n+1}},\\
  (\mathbf{b}_n)_{(\beta_{n-1},i_n,i_{n+1},\beta_{n+1})}
   & := \sum_{\beta_n}B_{\beta_{n-1}i_n\beta_n}B_{\beta_{n}i_{n+1}\beta_{n+1}}.
\end{align}

\jj{These expressions have a direct tensor-network reading, illustrated in Figure~\ref{fig:forms}. The inner product $\langle\mathbf{v},\mathbf{w}\rangle$ is the full contraction of the two-layer network built from the tensors of $\mathbf{v}$ and $\mathbf{w}$ [Fig.~\ref{fig:forms}(a)]. Fixing attention on the active tensor and contracting all the others collapses the surrounding network into an effective \textit{environment} [Fig.~\ref{fig:forms}(b)], so that the form reduces to an ordinary scalar product between effective vectors [Fig.~\ref{fig:forms}(c)]. The expectation value $\langle\mathbf{v},H\mathbf{w}\rangle$ reduces analogously to a quadratic form, with $Q$ the effective matrix obtained by contracting the MPO $H$ against the surrounding tensors.}

With these tools, the distance between two MPS becomes a quadratic formula $\Vert{\mathbf{v}-\mathbf{w}}\Vert^2
\sim \mathbf{a}^\dagger_n Q_n\mathbf{a}_n - 2\mathrm{Re}(\mathbf{a}^\dagger_n\mathbf{f}_n) + \Vert{\mathbf{w}}\Vert^2$, over a single unknown object $\mathbf{a}_n$. By computing the optimal $\mathbf{a}_n$, we obtain a vector of numbers that can be optimally split into neighboring tensors that minimize the distance and the tensor size. By repeating this task iteratively over all pairs of sites $(n,n+1)$\jj{---a left-to-right and right-to-left traversal we refer to as a \textit{sweep}, following the DMRG literature}---the task of compressing MPS (Sect.~\ref{sec:simplification}) becomes a series of easily solved local quadratic problems that rapidly converge to a solution.

Experts in DMRG will recognize that these two abstractions, \texttt{AntilinearForm} and \texttt{QuadraticForm}, are the basic ingredients of a two-site DMRG algorithm~\cite{schollwock2011}\jj{: the vectors $\mathbf{f}_n$ and matrices $Q_n$ are precisely the left and right \textit{environments} that contract everything outside the active pair $(n,n+1)$, here computed and cached as the sweep advances}. They are also the fundamental components of the MPS-based generalizations for tensor \jj{compression} (Sect.~\ref{sec:simplification}), linear combinations (Sect.~\ref{sec:addition}), eigenvalue search (Sect.~\ref{sec:DMRG}), matrix inversion (Sect.~\ref{sec:DMRG-solve}), etc.

\section{Linear Algebra Package}
\label{sec:LAPACK}
The previous section described a complete algebra associated with a vector space's elements and linear operators acting on them. Using this algebra, we can now address higher-level tasks, such as solving a linear equation, computing eigenvalues, and implementing the Fourier transform. This is what we call a true linear algebra package, in analogy with the well-known LAPACK standard for traditional algebra. These intermediate-level algorithms will be the basis for actual mathematical applications in the solution of differential equations, interpolation, and other real-world applications from Sections~\ref{sec:function-representation} to \ref{sec:integration}.

\subsection{Eigenvalue Search}
\label{sec:eigenvalue}
This task refers to solving the problem of computing both the value $\lambda$ and the vector $\mathbf{v}$ such that
\begin{equation}
  H \mathbf{v} = \lambda \mathbf{v},
\end{equation}
for some matrix product operator $H$. We enumerate now three different algorithms: the well-known DMRG method from condensed matter physics (Sect.~\ref{sec:DMRG}), the power method for very large or very small eigenvalues (Sect.~\ref{sec:power}), and the finite-precision version of the Arnoldi algorithm (Sect.~\ref{sec:Arnoldi}, Ref.~\cite{garcia-molina2024}). Each algorithm has its pros and cons, and they have all been rather exhaustively compared in Ref.~\cite{garcia-molina2024}. There is a fourth family of methods, known as imaginary time evolution, but, given their little reliability (c.f. Ref.~\cite{garcia-molina2024}), they are not documented here.

\subsubsection{DMRG}
\label{sec:DMRG}
The Density Matrix Renormalization Group or DMRG is an algorithm developed by Steven White~\cite{white1992, white1993} to compute the ground-state of 1D many-body Hamiltonians. It works by inspecting pairs of neighboring quantum objects and writing down a smaller effective problem that describes their direct interactions plus all the influence from surrounding quantum systems. This smaller problem is solved and used to analyze the state of another pair of neighboring quantum systems, repeating this process until the algorithm converges and an approximate representation of the entire quantum state is obtained~\cite {schollwock2005, schollwock2011}.

Nowadays, we know that DMRG implicitly builds an MPS tensor network~\cite{ostlund1995, dukelsky1998}. We also understand and identify the effective model as an approximation of the energy using a quadratic form and pairs of neighboring tensors~\cite{schollwock2011}, as explained in Sect.~\ref{sec:forms}. Specifically, we use
\begin{align}
  \langle\mathbf{v}, H\mathbf{v}\rangle &\sim \mathbf{a}_n^\dagger H_n\mathbf{a}_n,\\
  \langle\mathbf{v}, \mathbf{v}\rangle&\sim \mathbf{a}_n^\dagger N_n \mathbf{a}_n,
\end{align}
to exchange the problem of finding an eigenvector $H \mathbf{v}= E\mathbf{v}$ with the problem of computing the eigenvalue and local tensors in $H_n \mathbf{a}_n = E N_n \mathbf{a}_n.$

SeeMPS implements the two-site DMRG algorithm in its MPS generalization as a single function \rtt{algorithms/dmrg}{dmrg} that accepts as input an MPO $H$, a possible guess for the eigenstate $\mathbf{v}$, and some \jj{compression} strategies, and searches for the minimum eigenvalue
\begin{equation}
  E = \mathrm{min}_\mathbf{v} \frac{\langle\mathbf{v},H\mathbf{v}\rangle}{\langle\mathbf{v},\mathbf{v}\rangle},
\end{equation}
iterating until convergence. Note that, while it is possible to do so, SeeMPS does not yet implement targeting of other eigenvalues or symmetries.

\subsubsection{Gradient Descent}
\label{sec:descent}

Gradient-descent methods provide a simple approach to approximate
extremal eigenvalues of an operator by iteratively improving a trial
state. For finding the minimal eigenvalue, these methods update the
approximation by a displacement along the opposite direction of the
energy gradient, corresponding to the fastest local decrease.

Starting from an initial MPS $\mathbf{v}_k$, one iteration of the SeeMPS \rtt{algorithms/gradient_descent}{gradient\_descent} algorithm consists of
applying the operator $H$ to the state and updating it along the
corresponding descent direction,
\begin{equation}
  \mathbf{v}_{k+1} =
  \mathbf{v}_k + \Delta\beta \frac{\delta E}{\delta \psi}, \quad \mbox{ with } \frac{\delta E}{\delta\psi} = (H - \braket{H} \mathbb{I}) \mathbf{v}_k,
\end{equation}
where $\langle H \rangle = \langle \mathbf{v}_k, H \mathbf{v}_k \rangle$
and $\Delta\beta<0$ is a scalar step size. \jj{The direction $(H-\langle H\rangle\mathbb{I})\mathbf{v}_k$ is the gradient of the energy expectation value $\langle\mathbf{v},H\mathbf{v}\rangle$ under the unit-norm constraint $\langle\mathbf{v},\mathbf{v}\rangle=1$, the shift by $\langle H\rangle$ removing the component along $\mathbf{v}_k$; accordingly, the state is renormalized after each update.} MPS-based machine learning
methods~\cite{Gorodetsky2018b, Wang2020, Barratt2022} already use
gradient-descent techniques, where $\Delta\beta$ is usually referred to
as the learning rate.

While a fixed step size requires careful calibration, it is possible to
determine an optimal update parameter by minimizing the energy along the
descent direction. Introducing the shifted operator
$H' = H - \langle H \rangle \mathbb{I}$, the optimal value of
$\Delta\beta$ can be expressed in terms of the expectation values
$\langle H'^2 \rangle$ and $\langle H'^3 \rangle$, and evaluated using a
small number of MPO--MPS applications.

SeeMPS implements this gradient-descent strategy as a lightweight
baseline method, useful for exploratory calculations or as an
initialization for more robust algorithms.

\subsubsection{Power Method}
\label{sec:power}

The power method, or power iteration, is an algorithm to approximate
the \jj{eigenvalue of largest magnitude} of an operator and its corresponding eigenvector.
Starting from an initial MPS $\mathbf{v}_0$, each iteration applies the
operator and normalizes the result,
\begin{equation}
  \mathbf{v}_{k+1} = \frac{H \mathbf{v}_k}{\| H \mathbf{v}_k \|}.
\end{equation}

Convergence \jj{to a vector within the dominant eigenspace requires that the initial state has nonzero overlap with it; the individual eigenvector is determined only when the largest-magnitude eigenvalue is non-degenerate}. The convergence rate depends on the spectral gap
between the first two eigenvalues, so a larger gap generally yields
faster convergence. In \rtt{api/function/seemps.optimization.power_method}{power\_method}, each iteration requires a single MPO-MPS
application and normalization, with no additional linear combinations.

To target the smallest eigenvalue of a non-negative operator, the method
can be combined with operator inversion, using $(H-\epsilon)^{-1}$.
In this case, each iteration involves solving a linear system, which in
SeeMPS is done with a conjugate-gradient solver. \jj{The cost is then set by
two nested loops: each outer power iteration solves one shift-invert linear
system, whose own cost is that of the inner CGS iterations (Sect.~\ref{sec:CGS});
the remaining per-iteration bookkeeping is one MPO--MPS product and two MPS
linear combinations.} This makes the method more costly than the standard
power iteration, but still conceptually straightforward.

\subsubsection{Arnoldi with Restart}
\label{sec:Arnoldi}

The Arnoldi iteration generalizes the power method by operating in a
Krylov subspace of order $L$,
\begin{equation}
  \mathcal{K}_L(\mathbf{v}_k,H) = \mathrm{span}\{\mathbf{v}_k, H\mathbf{v}_k, \dots, H^{L-1}\mathbf{v}_k\},
\end{equation}
generated by repeated applications of $H$ to the current approximation
$\mathbf{v}_k$. Each iteration updates the state as a linear combination
of these basis vectors,
\begin{equation}
  \mathbf{v}_{k+1} = \sum_{m=0}^{L-1} v_m H^m \mathbf{v}_k,
\end{equation}
where the coefficients $\mathbf{v} = (v_0, \dots, v_{L-1})^T$ are obtained by solving the small generalized eigenvalue problem $ A \mathbf{v} = \lambda N \mathbf{v}$, with matrices $A$ and $N$ representing the projections of $H$ and the identity in the Krylov basis. This problem can be solved analytically and numerically, providing both a new estimate of the eigenvalue and eigenvector. The normalization matrix $N$ is relevant for the MPS implementation since the truncation error affects the orthogonality of the basis. \jj{For this reason we keep the generalized formulation rather than a plain Lanczos recurrence: under MPS truncation the Krylov vectors are not numerically orthonormal, so the overlap matrix $N$ cannot be assumed to be the identity. For a Hermitian $H$---the case handled by \texttt{arnoldi\_eigh}---this amounts to a truncation-robust Lanczos iteration that solves the generalized problem $A\mathbf{v}=\lambda N\mathbf{v}$ explicitly.}

Note that for $L= 1$, this method is equivalent to the power method, and for $L=2$, to the gradient descent method. In general, increasing the number of Krylov vectors may accelerate convergence, but numerical rounding errors (and finite MPS precision) can lead to numerical instabilities and a bad conditioning of the $N$ matrix. For this reason, in \rtt{algorithms/arnoldi}{arnoldi\_eigh} the number of Krylov vectors is typically restricted to a user provided value and, when this size is reached or $N$ runs the risk to become ill-conditioned, the subspace can be explicitly restarted with the new approximation $\mathbf{v}_{k+1}$ as the starting point. Furthermore, optional memory factors may be included in the eigenvalue search to improve convergence \cite{Pollock2021}.

\jj{As a rule of thumb for choosing among these solvers: DMRG (Sect.~\ref{sec:DMRG}) is a versatile, well-rounded default that exploits the variational MPS structure directly and, while originally devised for ground states of many-body Hamiltonians, tends to perform well across a broad range of problems, including the eigenvalue tasks arising in PDEs; the Arnoldi iteration is a robust general-purpose Krylov alternative; the power method is the cheapest per iteration and is best reserved for extremal eigenvalues with a large spectral gap; and gradient descent is a lightweight baseline or initializer. A systematic empirical comparison of these algorithms is reported in Ref.~\cite{garcia-molina2024}.}

\subsection{Linear Equation Solvers}
\label{sec:linear-equation}
The next problem we address is the inversion or solution of a linear equation
\begin{equation}
  \label{eq:linear-system}
  U \mathbf{x} = \mathbf{b}
\end{equation}
where $U$ and $\mathbf{b}$ are given MPO and MPS, and our goal is to find an MPS $\mathbf{x}$ that approximates the solution. SeeMPS offers three families of solvers, all of which essentially handle the minimization $\mathrm{argmin}_\mathbf{x}\Vert{U\mathbf{x}-\mathbf{b}}\Vert^2$ using either high-level iterations known from numerical analysis (Sects.~\ref{sec:CGS} and \ref{sec:GMRES}), or the optimization DMRG algorithm introduced before (Sect.~\ref{sec:DMRG-solve}). \jj{In practice, CGS is the natural choice for Hermitian or symmetric $U$, while BICGS and GMRES handle general non-Hermitian operators; the DMRG-based solver, in turn, is a versatile all-round option that often performs well---including for the linear systems and PDE problems addressed here---by exploiting the variational MPS structure directly.}

\subsubsection{CGS and BICGS}
\label{sec:CGS}
Using the BLAS from Sect.~\ref{sec:BLAS}, we can write down various linear algebra routines, starting with the celebrated conjugate gradient solver (\rtt{api/function/seemps.solve.cgs_solve}{cgs\_solve}) for symmetric and Hermitian MPOs, and its sibling, the bi-conjugate gradient solver (\rtt{api/function/seemps.solve.bicgs_solve}{bicgs\_solve}) for generic linear operators. SeeMPS implements both iterative methods, using standard formulations~\cite{wiki:cgs2025, wiki:bicgs2025}.

\begin{figure}[t]
\begin{pythoncode}
def cgs_solve(A: MPS, b: MPS, guess: MPS,
               strategy: Strategy):
  x = simplify(guess)
  r = b - A @ x
  p = simplify(r, strategy)
  residual = r.norm()
  for i in range(maxiter):
      if residual < tolerance * normb:
          break
      a = (residual**2) / A.expectation(p)
      x = simplify(x - a*p, strategy)
      r = b - A @ x
      residual, oldres = r.norm(), residual
      p = r + (residual / oldres) * p
      p = simplify(p, strategy)
  return x, abs(residual)
\end{pythoncode}
  \caption{Pseudocode implementation of the conjugate gradient solver algorithm with MPS and MPO.}
  \label{fig:cgs}
\end{figure}
To illustrate the power of working with the MPS BLAS, in Fig.~\ref{fig:cgs} we offer here the code for \texttt{cgs\_solve}, stripped from some optional functionality.
Note how the code is almost indistinguishable from standard implementations of CGS using Python and NumPy, except for the implicit \texttt{simplify} operations that keep the size of the MPS in check (See Sect. \ref{sec:simplification}).

\subsubsection{Krylov Methods (GMRES)}
\label{sec:GMRES}
The SeeMPS library implements (\rtt{api/function/seemps.solve.gmres\_solve}{gmres\_solve}) the Generalized Minimal Residual (GMRES) method for solving general non-Hermitian linear systems $U\mathbf{x} = \mathbf{b}$. This iterative solver builds an approximate solution $\mathbf{x}_m$ within the Krylov subspace of dimension $m$ generated by the initial residual $\mathbf{r}_0 = \mathbf{b} - U\mathbf{x_0}$,
\begin{equation}
\mathcal{K}_m(\mathbf{r}_0, U) = \mathrm{span}\{\mathbf{r}_0, U\mathbf{r}_0, \ldots, U^{m-1}\mathbf{r}_0\}.
\end{equation}
The method approximates $\mathbf{x}_m$ by minimizing the residual norm $\|\mathbf{b} - U\mathbf{x}_m\|$ over this subspace.

An orthonormal basis $\{\mathbf{v}_1, \ldots, \mathbf{v}_m\}$ for $\mathcal{K}_m$ is built via the Arnoldi iteration with modified Gram-Schmidt orthogonalization. Since each iteration involves MPO-MPS contractions and linear combinations that tend to increase the bond dimension, \jj{compression} steps are applied to keep the MPS representation tractable.

Once the basis is built, the residual minimization reduces to a small $(m+1) \times m$ least-squares problem, which is solved using NumPy to obtain the optimal coefficients $\mathbf{y}$. The approximate solution is then constructed as
\begin{equation}
  \mathbf{x}_m = \mathbf{x}_0 + \sum_{j=1}^{m} y_j \mathbf{v}_j,
\end{equation}
where the linear combination is accumulated efficiently as an \texttt{MPSSum} and variationally compressed.

The implementation supports restarted GMRES, where, upon reaching the maximum subspace dimension, the current approximation $\mathbf{x}_m$ becomes the initial guess for a fresh Krylov subspace.

\subsubsection{\jj{DMRG as a Linear System Solver}}
\label{sec:DMRG-solve}
The DMRG algorithm can also be used to solve the linear system~\eqref{eq:linear-system}. Our implementation is philosophically equivalent to the vector correction method from \jj{Refs.~\cite{ramasesha1997, kuhner1999}}. Using the quadratic and antilinear forms from Sect.~\ref{sec:forms}, we rewrite the complete system as a projected version acting on pairs of tensors $(n,n+1)$ of the unknown
\begin{equation}
  \label{eq:intermediate-equation}
  U_{n} \mathbf{a}_n - \mathbf{b}_n = 0,
\end{equation}
This system of equations can be solved using any standard algorithm (e.g. SciPy's CGS, BICGS, etc), to obtain new estimates for a pair of tensors $\mathbf{A}^{(n)}, \mathbf{A}^{(n+1)}$, used to update the estimate $\mathbf{x}$. By repeating this step many times, back and forth along the tensor train, the algorithm quickly converges to the optimal solutions, within the given constraints of tolerance, truncation limits, and bond dimension.

\subsection{Discrete Fourier Transform}
\label{sec:QFT}
We include in this sublibrary the MPS equivalent of the Fast Fourier Transform (FFT), an algorithm that transforms vectors in a space of dimension $2^n$ as
\begin{equation} \label{eq:QFT}
  \mathbf{e}_{i}\xrightarrow{\hat{\mathcal{F}}}\frac{1}{\sqrt{2^n}} \sum_{j=0}^{2^n-1} e^{-\jj{\mathrm{i}} 2 \pi ij / 2^n} \mathbf{e}_j.
\end{equation}
In SeeMPS, the Fourier transform is implemented as a sequence of unitary transformations, $\mathcal{F} = F_n F_{n-1}\cdots F_1$, that mimic layers of a Quantum Fourier Transform circuit acting on $n$ qubits~\cite{coppersmith2002}, up to the final qubit swap.

Specifically, SeeMPS assumes that the Fourier transform acts on an MPS composed of $n$ two-dimensional objects or qubits. The total algorithm is encoded as an \texttt{MPOList} with $n$ MPOs implementing Hadamard gates and conditional rotations, with exact tensors of size $2^4$
\begin{align}
B^{s'_n, s_n}_{n, 0, 0} &= \delta_{s'_n, s_n}, \quad n < i, \\
B^{s'_i, s_i}_{i, 0, s_i} &= H_{s'_i, s_i}, \quad \text{$i$-th qubit}, \\
B^{s'_j, s_j}_{j, s_i, s_i} &= \exp \left[ \frac{\jj{\mathrm{i}} 2\pi}{2^{j-i}} s_j s_i \right] \delta_{s'_j, s_j}, \quad j > i.
\end{align}

The library provides the functions \rtt{api/function/seemps.qft.qft}{qft} and \rtt{api/function/seemps.qft.iqft}{iqft} to do one-shot transforms on a given vector. It also offers constructors of MPOs that can be reused, for the direct \rtt{api/function/seemps.qft.qft_mpo}{qft\_mpo} and inverse transforms \rtt{api/function/seemps.qft.iqft_mpo}{iqft\_mpo}, and for partial transforms of a multidimensional tensor encoded in the quantum register (\rtt{api/function/seemps.qft.qft_nd_mpo}{qft\_nd\_mpo}, \rtt{api/function/seemps.qft.iqft_nd_mpo}{iqft\_nd\_mpo}).

The function \rtt{api/function/seemps.qft.qft_flip}{qft\_flip} implements the qubit reversal that makes \texttt{qft\_flip(qft(f))} equivalent to the FFT of the vector version of \texttt{f}. In both cases, negative frequencies are placed in the upper part of the quantum register, in what is known as two's complement notation. However, whenever possible, we try to avoid the \texttt{qft\_flip} operation, which is costly in the MPS representation.

\jj{What is} more important is the fact that the Fourier transform does not significantly change the amount of correlations and the bond dimension of an MPS. This was first reported for discrete encodings of bandwidth-limited functions~\cite{garcia-ripoll2021}, and it was later confirmed in more general scenarios, introducing approximate, low-rank versions of the QFT MPO that are very efficient~\cite{chen2023}. Either approach allows us to benefit from the exponential convergence of Fourier-based techniques in applications such as differentiation (Sect.~\ref{sec:differentiation}) and interpolation~\cite{garcia-molina2022}, as well as exponential speedups in certain applications where the FFT would be required~\cite{gidi2025, garcia-ripoll2021}.

\section{Function Representation and Encoding}
\label{sec:function-representation}
In this section, we discuss the encoding of functions as MPS or tensor trains, together with the tools to compute those encodings. The core idea is that we rely on the discretization of a multidimensional function $f(x_1,x_2,\ldots, x_{\jj{M}})$ \jj{of $M$ variables} on a grid with $\mathcal{O}(2^{\jj{M}n})$ points, using an MPS with $\jj{Mn}$ tensors to achieve an exponential compression $\mathcal{O}(\jj{Mn}\chi^2)$ (Sect.~\ref{sec:loading}). Naturally, this compression does not make a lot of sense if the computation of those tensors has an exponential cost, too. For that reason, we discuss in Sects.~\ref{sec:direct} to \ref{sec:complementary} a large variety of \textit{function loading} techniques that compute the MPS representation using either a computational description or efficient sampling techniques.

\subsection{Function Quantization}
\label{sec:loading}
\begin{figure}[t]
  \centering
  \includegraphics[width=0.9\linewidth]{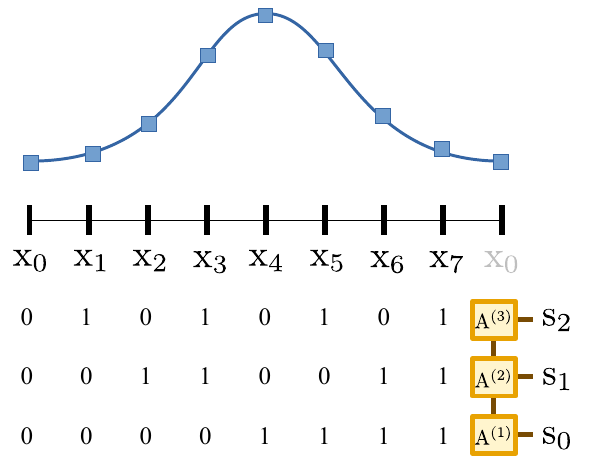}
  \caption{Discretization and encoding of a periodic function on a grid with 8 points, using an MPS with three qubits.}
  \label{fig:quantization}
\end{figure}
Let us imagine a one-dimensional function $f(x)$, defined over an interval $[a,b)$ that is sampled on a uniform grid of $2^n$ points, as sketched in Fig.~\ref{fig:quantization}. We can establish two mappings from $2^n$ integer values to the coordinates and the values of that function:
\begin{align}
  i \in\{0,1,\ldots,2^n-1\}
  &\to x_i^{(n)} = a + i \Delta{x}^{(n)},\\
  &\to v_i := f(x_i^{(n)}),
  \label{eq:position-quantization}
\end{align}
with the uniform spacing $\Delta x^{(n)}=(b-a)/2^n$.

As explained in Sect.~\ref{sec:vector}, the vector of values $v_i$ admits a decomposition or \textit{quantization} using a set of $n$ binary digits $s_i\in\{0,1\}$ or \textit{qubits} in the quantum notation
\begin{align}
  \label{eq:binary}
  i &= 2^{n-1}s_1 + 2^{n-2}s_2 + \ldots + s_n,\\
  v_i &= v_{s_1s_2\ldots s_{n}},
\end{align}
to obtain a tensor $v_{s_1s_2\ldots s_n}$ that can then be compressed using an MPS or tensor train decomposition
\begin{equation}
  \jg{
  f(x_i^{(n)}) \to \sum_{\lbrace \alpha \rbrace}
  (A_{\alpha_1}^{s_1}A_{\alpha_{1},\alpha_2}^{s_2}\dots A_{\alpha_{n-1}}^{s_{n}}).
  }
\end{equation}
Each tensor $A_{\alpha_{k-1},\alpha_{k}}^{s_k}\in\mathbb{C}^{2\times \chi_{k-1}\times \chi_{k}}$ has a bounded size, with dimensions $\chi_{k-1},\chi_k$ that depend on the entanglement content. Provided the bond dimensions remain bounded, the storage cost scales as $\mathcal{O}(2n\,\chi^2)$, compared to $\mathcal{O}(2^n)$ for a full grid representation. This is verified for broad classes of functions as discussed in Ref.~\cite{jobst2024}.

The idea we have just presented can be traced back to a combination of Zalka, Grover, and Rudolph's encoding of functions in quantum registers~\cite{zalka1998, grover2002}, with the realization that MPS can efficiently compress some instances of those quantum states. This \textit{quantum inspired} methodology has been rediscovered multiple times. In the physics domain,  it has been introduced as a multigrid technique to solve nonlinear Schrödinger equations\ \cite{lubasch2018}, turbulence problems~\cite{gourianov2022}, and also a broad family of function manipulation and equation solution techniques~\cite{garcia-ripoll2021}. This technique, however, was first introduced in the field of applied mathematics by Oseledets~\cite{oseledets2010}, in what is known as the \textit{quantized tensor train (QTT) algorithm}~\cite{khoromskij2010, khoromskij2011}.

SeeMPS extends this technique to the encoding of multidimensional functions $f(\mathbf{x})=f(x_1,\dots,x_{\jj{M}})$. As before, we label the coordinates over grids that are defined for each of the $\jj{M}$ axes  $x_{i_m}\in[a^{m},b^m)$, and the corresponding integer labels $i_m$ are quantized using $n_m$ qubits each $\{s_k^m\}$, resulting in an MPS with a total of $N=\sum_{\jj{m=1}}^{\jj{M}} n_m$ tensors. As explained in Sect.~\ref{sec:outer}, there are different orders in which we can arrange the qubits in the MPS, and those orders will affect the entanglement structure and size of the tensors we have to use~\cite{garcia-ripoll2021}. SeeMPS currently supports the  coordinate-major (``A'') and scale-interleaved (``B'') orders which, for a two-dimensional function $f(x_1,x_2)$ with 3 qubits for each coordinate, results in two maps
\jg{
\begin{align}
  \text{Order A:} (i_1,i_2) &\rightarrow (s_1^1,  s_2^1,  s_3^1,  s_1^2,  s_2^2,  s_3^2), \\
  \text{Order B:} (i_1,i_2) &\rightarrow (s_1^1,  s_1^2,  s_2^1,  s_2^2,  s_3^1,  s_3^2),
\end{align}
}
but certain algorithms, such as TCI, support more general orders.

More generally, functions discretized on non-equispaced or irregular grids can also be loaded in MPS form. While Eq.~\eqref{eq:position-quantization} introduces the samples $v_i = f(x_i^{(n)})$ using a uniform grid for $x_i^{(n)}$, an equivalent construction applies to non-uniform discretizations by identifying the uniformly spaced variable with the index itself, $x_i^{(n)}=i$, and absorbing the nonlinear mapping to physical space into the sampled values $v_i$. As an example, a function sampled at the Chebyshev--Lobatto nodes $x_j = \cos\left( \frac{j\pi}{N} \right)$ can be represented as $v_j = f(\cos \tfrac{j \pi}{N})$, where the index $j$ remains uniformly spaced and admits a binary quantization. Such irregular discretizations yield spectral function representations that enable high-accuracy algorithms, including spectral interpolation and fast quadrature rules (see Sec.~\ref{sec:integration}).

\subsection{Direct Construction and Polynomials}
\label{sec:direct}
Many simple functions admit closed-form QTT/MPS representations, where all tensor entries can be prescribed analytically~\cite{oseledets2013}. For instance, a linear combination of exponentials
\begin{equation}
  f(x) = \sum_{n=1}^{N}w_n \exp(k_n x),
\end{equation}
can be encoded in an MPS with bond dimension $\chi=N$. The function \rtt{api/function/seemps.analysis.factories.mps_sum_of_exponentials}{mps\_sum\_of\_exponentials} implements this construct, which is used in term by \rtt{api/function/seemps.analysis.factories.mps_exponential}{mps\_exponential}, \rtt{api/function/seemps.analysis.factories.mps_cos}{mps\_cos} and \rtt{api/function/seemps.analysis.factories.mps_sin}{mps\_sin} to encode one exponential and two trigonometric functions.

A richer, fully explicit, and less obvious example is given by one-dimensional polynomials~\cite{lindsey2024, rodriguez-aldavero2025}. We may construct any degree-$\jj{q}$ polynomial
\begin{equation}
   f_{\jj{q}}(x)=\sum_{i=0}^{\jj{q}} p_i x^i,
\end{equation}
defined on a discretized interval $x\in[a,b]$, using an MPS with bond dimension $\chi\leq \jj{q}+1$ (in practice, often smaller, if one uses the compression routines). This algorithm is implemented in \rtt{api/function/seemps.analysis.polynomials.mps_from_polynomial}{mps\_from\_polynomial}, a function that can either take the coefficients $\{p_i\}$ or convert a NumPy representation of a polynomial to MPS form.

\jr{Other useful encodings include the coordinate MPS \rtt{api/function/seemps.analysis.factories.mps_interval}{mps\_interval}, which represents an \texttt{Interval} object as an MPS. For a regular interval, its entries coincide with the equispaced grid points $x_i = a + i \Delta x$, yielding the MPS analogue of \texttt{numpy.linspace}. This coordinate MPS is standard argument passed to the function-composition routines used throughout Refs.~\cite{garcia-ripoll2021, garcia-molina2024, garcia-molina2025}. Another useful example is the Heaviside step function $\Theta(x)$, implemented as \rtt{api/function/seemps.analysis.factories.mps_heaviside}{mps\_heaviside}. Both constructions have bond dimension $\chi = 2$.} However, equally important is the fact that we can create new functions by composing other MPS via multiplication (the element-wise product from Sect.~\ref{sec:inner-product}), scaling, and addition (Sect.~\ref{sec:addition}) of previously encoded functions, as well as encode multivariate functions via outer operations (Sect.~\ref{sec:outer}). For instance, $|x|$ may be constructed as $x (\Theta(x) -  \Theta(-x))$, using those operations, to arrive at an MPS with bond dimension $\chi=2$ or 3, depending on the interval.

\subsection{Orthogonal Polynomial Expansions}
\label{sec:orthogonal-polynomials}
Smooth functions can often be accurately approximated by truncated expansions over a complete polynomial basis $\{P_k(x)\}$,
\begin{equation}
  \label{eq:polynomial}
  f(x) \approx \sum_{k=0}^{\jj{q}} c_k P_k(x).
\end{equation}
This enables both the direct encoding of a function $f(x)$ and the composition $f\circ g$ of $f$ with another function $g$ already represented as an MPS~\cite{rodriguez-aldavero2025}. Specifically, given an MPS $\mathbf{v}$ encoding the function $g(x)$, we can compute the encoding $\mathbf{w}$ of the composed function through the expansion
\begin{equation}
  (f\circ g)(x)=f(g(x)) \to \mathbf{w} = \sum_k c_k P_k(\mathbf{v}).
\end{equation}
The same technique applies straightforwardly to operator-valued functions, allowing the evaluation of $f(O)$ for an operator $O$ represented as an MPO.

SeeMPS provides a unified framework for such polynomial expansions, with implementations tailored to the choice of polynomial basis. The class \rtt{api/class/seemps.analysis.expansion.PowerExpansion}{PowerExpansion} implements expansions in the monomial basis $P_k(x)=x^k$, for which the user must explicitly provide the coefficients $\{c_k\}$.

More generally, SeeMPS supports expansions in arbitrary orthogonal polynomial bases relying on three-term recurrence relations
\begin{equation}
  \label{eq:Clenshaw}
P_{k+1}(x) = (\alpha_k x + \beta_k) P_k(x) - \gamma_k P_{k-1}(x)
\end{equation}
which are evaluated using numerically stable Clenshaw formulas~\cite{clenshaw1955}. In this setting, the user provides the target function $f$ together with an initial MPS or MPO encoding the argument. The methods \rtt{api/class/seemps.analysis.expansion.PolynomialExpansion}{to\_mps} and \rtt{api/class/seemps.analysis.expansion.PolynomialExpansion}{to\_mpo} then construct the corresponding MPS/MPO compositions. Internally, the expansion coefficients $\{c_k\}$ are computed by projection or collocation onto the selected polynomial basis.

The Chebyshev and Legendre polynomial families are implemented by default via the classes \rtt{api/class/seemps.analysis.expansion.ChebyshevExpansion}{\path{ChebyshevExpansion}} and \rtt{api/class/seemps.analysis.expansion.LegendreExpansion}{\path{LegendreExpansion}}. In particular, the use of Chebyshev polynomials yields an approximation framework analogous to MATLAB's ``ChebFun'' package~\cite{driscoll2014}, but formulated entirely within the MPS/MPO formalism~\cite{holzner2011, halimeh2015, rodriguez-aldavero2025}.

\begin{figure}[t]
\begin{pythoncode}
import numpy as np
from seemps.state import mps_tensor_sum
from seemps.analysis.mesh import QuantizedInterval
from seemps.analysis.factories import mps_interval
from seemps.analysis.expansion import ChebyshevExpansion

(a, b), N = (-1, 1), 10
interval = QuantizedInterval(a, b, N)
mps_x = mps_interval(interval)
mps_xy = mps_tensor_sum([mps_x] * 2)
f = lambda x: np.exp(x)
expansion = ChebyshevExpansion.project(f)
mps_f = expansion.to_mps(argument=mps_xy)
\end{pythoncode}
  \caption{Pseudocode to encode a multivariate function $f(x, y) = e^{x + y}$, $x, y \in [-1, 1]$ in MPS with a Chebyshev expansion. \jj{This separable example is chosen to illustrate the Chebyshev API; being a product $e^x e^y$, it can in fact be represented directly at bond dimension one.}}
  \label{fig:cheb}
\end{figure}

The applicability of this technique is limited by the regularity of the target function. Highly differentiable functions present favorable convergence rates,  while functions with discontinuities or sharp features---such as Heaviside functions---are poorly approximated by polynomial expansions, requiring prohibitively large expansion orders.

\subsection{Function Interpolation}
\label{sec:interpolation}

Function interpolation estimates the values of a discretized function between the points of its original grid. Standard numerical analysis interpolation techniques are provided in the SeeMPS library enabled by the MPS finite-precision BLAS.

\subsubsection{Finite Differences Interpolation}
\label{sec:fd_interpolation}
The finite differences interpolant uses piecewise linear or polynomial interpolation to approximate new points of the $(n+1)$-qubit grid via a Taylor expansion
\begin{equation}
  f(x_s^{(n)} + \varepsilon) = f(x_s^{(n)}) + \varepsilon \, \partial_x f(x_s^{(n)})
  + O((\Delta x^{(n)})^2),
\end{equation}
with $\varepsilon \in [-\Delta x^{(n)}, \Delta x^{(n)}].$ Setting $\varepsilon = \Delta x^{(n)}/2$ yields a second-order
finite-difference interpolant,
\begin{equation}
  f\left(x_s^{(n)} + \frac{\Delta x^{(n)}}{2}\right) \approx
  f(x_s^{(n)}) + \frac{f(x_s^{(n)} + \Delta x^{(n)}) - f(x_s^{(n)})}{2}.
\end{equation}
The function \rtt{api/function/seemps.analysis.interpolation.finite_differences_interpolation}{finite\_differences\_interpolation}, implements this algorithm (and higher order versions) using a collection of displacement operators $D_k$ that is efficiently encoded as an MPO, and contracted with the function to produce a single MPS. Further details can be found in Sect.~\ref{sec:finite-differences}.

\subsubsection{Fourier Interpolation}
\label{sec:fourier_interpolation}

Fourier interpolation is particularly advantageous for bandwidth-limited functions, offering exponentially small errors in the number of qubits~\cite{Monro1979, garcia-ripoll2021}. According to the
Nyquist-Shannon theorem~\cite{Nyquist1928, Shannon1949}, a function
with maximum bandwidth $p_\text{max}$ can be reconstructed exactly
from samples taken at a rate $p_l \ge 2 p_\text{max}$. For a function
with domain sizes $L_x$ and $L_p$ in position and momentum space,
respectively, this requires
\begin{equation}
  \Delta x^{(n)} \le \frac{2\pi}{L_p}, \quad
  \Delta p^{(n)} \le \frac{2\pi}{L_x}.
\end{equation}

In this framework, the function is transformed to momentum space, extended by zero-padding outside the band\-width-limited region, and then transformed back to position space. Formally, the interpolated function on a finer $(n+m)$-qubit grid is defined as
\begin{align}
  \label{eq:interpolation}
  f^{(n+m)}(x) = \mathcal{F}^{-1} \, \mathcal{P} \, \big( \mathcal{F} f^{(n)}(x) \big)
  =: \hat{U}_\mathrm{int}^{\,n,m} f^{(n)}(x),
\end{align}
where $\mathcal{F}$ and $\mathcal{F}^{-1}$ are the discrete Fourier and inverse Fourier transforms, and $\mathcal{P}$ is the zero-padding operator that symmetrically embeds the original Fourier components within the larger momentum vector. The operator $\hat{U}_\mathrm{int}^{n,m}$ encodes the full interpolation from the $n$-qubit to the $(n+m)$-qubit representation.

The SeeMPS function \rtt{api/function/seemps.analysis.interpolation.fourier\_interpolation}{fourier\_interpolation} implements this method, where the minimum number of tensors required for an accurate representation follows the Nyquist criterion,
$n_\text{min} \jg{\ge} \log_2 \frac{L_p L_x}{2\pi}.$

\subsection{Tensor Cross-Interpolation (TCI)}
\label{sec:TCI}
Tensor cross-interpolation (TCI) is a sampling-based technique for constructing MPS representations of multivariate functions by considering them as black-box objects to be probed along structured index patterns, without ever forming the full discretization tensor~\cite{oseledets2010a}. This makes TCI particularly well-suited for high-dimensional problems, where explicit discretization would be prohibitively expensive.

At its core, TCI generalizes the \emph{skeleton decomposition}~\cite{goreinov1997}, a low-rank matrix factorization in which a matrix $A$ is reconstructed from a subset of its rows and columns,
\begin{equation}
A = C \hat{A}^{-1} R,
\end{equation}
where $C$ and $R$ are formed from selected columns and rows of $A$, and $\hat{A}$ is their intersection, known as the \emph{pivot} submatrix. While the optimal selection of pivots is hard~\cite{goreinov2010}, practical algorithms rely on \emph{maxvol} heuristics that aim to maximizing the \emph{volume}
\begin{equation}
\operatorname{vol} \hat{A} := |\det \hat{A}|.
\end{equation}

TCI extends this idea to higher-order tensors by applying the skeleton decomposition to successive matrix unfoldings~\cite{oseledets2010a}. Given a tensor $A_{s_1\ldots s_L}$, the unfolding at position $k$ is defined as
\begin{equation}
A_{s_1\ldots s_L} \longrightarrow A_{s_{\le k} \ s_{>k}},
\end{equation}
where $s_{\le k}$ groups the first $k$ indices and $s_{>k}$ the remaining ones. Modern TCI algorithms iteratively refine an initial approximation by probing the underlying black-box function along structured index patterns, known as \emph{fibers}, and updating the MPS tensors \jj{(also called \emph{cores} in the tensor-train literature)} accordingly.

SeeMPS provides several TCI variants, accessible through a common interface \rtt{api/function/seemps.analysis.cross.cross_interpolation}{cross\_interpolation}. The function \rtt{api/function/seemps.analysis.cross.cross_maxvol}{cross\_maxvol} implements a rank-adaptive variant based on single-site rectangular-maxvol updates~\cite{sozykin2022, mikhalev2018}, the function \rtt{api/function/seemps.analysis.cross.cross_dmrg}{cross\_dmrg} implements a two-site variant with DMRG-like updates~\cite{savostyanov2011}, and \rtt{api/function/seemps.analysis.cross.cross_greedy}{cross\_greedy} provides an alternative approach based on greedy pivot selection~\cite{dolgov2020, nunez-fernandez2022, ritter2024}. In practice, we find the DMRG-based approach to be the most efficient, although it can be more susceptible to convergence to local minima, whereas the rectangular-maxvol approach is typically slower but more robust. Other variants, such as partial rank-revealing LU schemes~\cite{nunez-fernandez2025}, follow the same underlying principle: sample the target function at informative fibers and update the MPS cores accordingly.

The black-box abstraction allows SeeMPS to address a wide range of problems within a unified framework. The standard use case is encoding a multivariate function MPS via the \rtt{api/class/seemps.analysis.cross.BlackBoxLoadMPS}{BlackBoxLoadMPS} class. To support any MPS core arrangement, including serial and interleaved schemes (see Sec.~\ref{sec:loading}), this requires an explicit linear mapping between the MPS indices $\mathbf{s}$ and discretization coordinates $\mathbf{x}$, which can be computed using the function \rtt{api/function/seemps.analysis.mesh.mps_to_mesh_matrix}{mps\_to\_mesh\_matrix}. This decoupled approach enables encoding functions in arbitrary QTT core arrangements.

This approach extends naturally to more sophisticated scenarios, such as the encoding of vector- and tensor-valued fields by treating the additional components as extra tensor indices. Moreover, functions can be represented directly in spectral space by encoding the corresponding tensor of expansion coefficients—for example, in Fourier or Chebyshev bases—rather than in physical space~\cite{bigoni2016}. Such spectral encodings enable fast interpolation, efficient evaluation of derivatives, and compact representations that are particularly well suited for the numerical solution of partial differential equations within the MPS/MPO framework~\cite{adak2025}.

Additional use cases of TCI include function composition by evaluating a target function on one or more MPS inputs $\ket{g}$ that are treated as black-box oracles, via the \rtt{api/class/seemps.analysis.cross.BlackBoxComposeMPS}{\path{BlackBoxComposeMPS}} class. \jr{This same class can be used to apply MPS compression.} Also, the class \rtt{api/class/seemps.analysis.cross.BlackBoxLoadMPO}{\path{BlackBoxLoadMPO}} supports the construction of one-dimensional MPOs.

\begin{figure}[t]
\begin{pythoncode}
import numpy as np
from seemps.analysis.mesh import QuantizedInterval, Mesh, mps_to_mesh_matrix
from seemps.analysis.cross import BlackBoxLoadMPS, cross_interpolation, CrossStrategyDMRG

(a, b), N = (-1, 1), 10
interval = QuantizedInterval(a, b, N)
f = lambda x: np.exp(np.sum(x, axis=0))
mesh = Mesh([interval, interval])
M = mps_to_mesh_matrix([N, N]) # Map matrix
dims = [2] * 2 * N # Physical dimensions
bb = BlackBoxLoadMPS(f, mesh, M, dims)
strategy = CrossStrategyDMRG()
mps = cross_interpolation(strategy, bb).mps
\end{pythoncode}
  \caption{Pseudocode to encode a multivariate function $f(x, y) = e^{x + y}$, $x, y \in [-1, 1]$ in MPS using TCI-DMRG. \jj{The black box receives the sampled coordinates stacked along the first axis of the input array, so \texttt{np.sum(x, axis=0)} sums the per-coordinate contributions $x+y$ at each probed point.}}
  \label{fig:tci}
\end{figure}

\subsection{Complementary Techniques}
\label{sec:complementary}
Beyond polynomial expansions and TCI, SeeMPS incorporates several complementary algorithms from the literature that are particularly effective in specialized regimes. These methods are not intended as general replacements, but rather address scenarios in which standard approaches become inefficient or fail to converge.

\subsubsection{Multiscale Interpolative Constructions.}
Multiscale interpolation methods~\cite{lindsey2024, chen2026} construct MPS encodings by interpolating functions on hierarchically refined dyadic grids. In SeeMPS, they are included as efficient construction routines for smooth, low-dimensional functions, often outperforming TCI and polynomial expansions~\cite{rodriguez-aldavero2025}. These methods produce highly sparse MPS representations in which all cores are uniform except for a single core that encodes the function samples. Their applicability is limited to low-dimensional problems, as they rely on direct sampling and therefore require a number of function evaluations that grows exponentially with the dimension. Implementations are provided by \rtt{api/function/seemps.analysis.lagrange.mps_lagrange_chebyshev_basic}{\path{mps_lagrange_chebyshev_basic}}, with rank adaptation via \rtt{api/function/seemps.analysis.lagrange.mps_lagrange_chebyshev_rr}{\path{mps_lagrange_chebyshev_rr}} and local interpolation via \rtt{api/function/seemps.analysis.lagrange.mps_lagrange_chebyshev_lrr}{\path{mps_lagrange_chebyshev_lrr}}.

\subsubsection{Sketching Methods}
Sketching methods use randomized projections to efficiently identify low-rank structure in large tensor representations~\cite{hur2023}. In the context of function encoding, the TT-RSS (recursive sketching from samples) algorithm combines sketching with cross-interpolation ideas by organizing a fixed set of samples into tensor fibers and reconstructing TT cores via small least-square problems~\cite{pareja-monturiol2025}. In SeeMPS, this approach is implemented in the function~\rtt{api/function/seemps.analysis.sketching.tt_rss}{tt\_rss}. It is well-suited for encoding high-dimensional functions and probability densities, where TCI and polynomial expansions may fail to converge, trading exact reconstruction for scalable approximation from sampled data.

\subsubsection{Computation-tree Constructions}
Computation-tree methods represent multivariate functions with compositional or branching algebraic structure through an explicit user-defined evaluation graph, compressing intermediate results at each node to produce highly sparse MPS representations~\cite{ryzhakov2022}. In SeeMPS, these methods are implemented by the routines \rtt{api/function/seemps.analysis.comptree.mps_chain_tree}{mps\_chain\_tree} and \rtt{api/function/seemps.analysis.comptree.mps_binary_tree}{mps\_binary\_tree}. These methods remain efficient in high-dimensional settings when the target function exhibits suitable algebraic structure or is procedurally defined, even in the presence of sharp features, where polynomial expansions and TCI may struggle.

\section{Function Differentiation and Integration}
\label{sec:differentiation}
The SeeMPS library provides three different numerical differentiation strategies optimized for functions that are encoded in the MPS/TT format. The three techniques are finite differences schemes, spectral differentiation based on Fourier techniques, and Hermite Distributed Approximating Functionals (HDAF). Given a domain of application, all of them encode the differential operator as an MPO that can be applied one or multiple times, with implicit compression parameters to balance bond dimension growth and convergence rates.

\subsection{Finite Differences}
\label{sec:finite-differences}
SeeMPS offers \rtt{api/function/seemps.analysis.derivatives.finite_differences_mpo}{finite\_differences\_mpo}, a function to create finite-difference approximations to the first and second derivatives using both standard and higher-order optimized versions~\cite{snrd}. The result is always a linear combination of MPOs that involve the identity and displacement operators acting on the function itself.

For instance, the second-order centered approximation in a grid with $2^n$ points
\begin{equation}
\partial_x^2 f(x_i) \simeq
\frac{f(x_{i+1}) + f(x_{i-1})-2f(x_i)}{(\Delta x^{(n)})^2}
\end{equation}
is implemented as
\begin{equation}
  \partial_x^2 \to \frac{1}{(\Delta x^{(n)})^2}(D_{+1} + D_{-1} - 2 \mathbb{I}),
\end{equation}
with an operator that displaces the MPS vectors $D_{m}$ by $m$ units (see \rtt{api/function/seemps.register.mpo_weighted_shifts}{mpo\_weighted\_shifts}). That is, in periodic boundary conditions
\begin{equation}
  (D_m\mathbf{v})_i = v_{i + m \mbox{ mod } 2^n},
\end{equation}
or in open boundary conditions
\begin{equation}
  (D_m\mathbf{v})_i = \left\{\begin{array}{ll}
    v_{i + m}, & 0 \leq i + m < 2^n,\\
    \jj{0} & \mbox{else.}
  \end{array}\right.
\end{equation}

These MPOs have a bond dimension that is equal to the number of displacements ($\chi=3$ for the second derivative), and an error that decays algebraically with the grid step $\mathcal{O}((\Delta x)^2)$ and exponentially in the number of tensors  $\mathcal{O}(2^{-2n})$. Higher-order derivatives and higher-order stencils can be constructed using the same formalism.

\subsection{Fourier Methods}
\label{sec:Fourier-differentiation}
Function \rtt{api/function/seemps.analysis.derivatives.fourier_derivative_mpo}{fourier\_derivative\_mpo} provides spectral differentiation based on Fourier interpolation. This method achieves significantly higher accuracy than finite differences and allows us to compute \textit{any analytical function $G(\partial_x)$ of the differential operator}. Let us assume that $\mathbf{p}$ labels the MPS encodings of the vector of frequencies in the Fourier space, and $\mathcal{F}$ and $\mathcal{F}^{-1}$ are the MPOs associated to the direct and inverse Fourier transforms. The MPO encoding of a differential operator $G(\partial_x)$ can be written
\begin{equation}
G(\partial_x) \to
\mathcal{F}^{-1} G(\jj{\mathrm{i}}\mathbf{p}) \mathcal{F}.
\end{equation}
While the momentum operator $\hat{p}$ admits an MPO representation with bond dimension $\chi=2$, a naive MPO construction of the QFT leads to a bond dimension that grows linearly with the number of sites. Efficient implementations exploiting bit-reversal symmetry significantly reduce this cost, as demonstrated in Ref.~\cite{chen2023}.

For analytic and periodic functions, spectral differentiation achieves exponential convergence $\mathcal{O}(e^{-r\jj{2^n}})$ \jj{in the number of grid points $2^n$, which amounts to} a super-exponential convergence with the number of qubits $n$ in the tensor decomposition. However, when the function is not periodic, or the interval size does not match the period, the Fourier expansion may lead to localized Gibbs oscillations that decay exponentially with the number of qubits\jj{,} $\mathcal{O}(2^{-n\alpha})$.

\subsection{\jj{Hermite Distributed Approximating Functionals (HDAF)}}
\label{sec:HDAF}
The Hermite Distributed Approximating Functionals~\cite{hoffman1991} are well-tempered approximations of the Dirac-delta distribution
\begin{equation}
  \label{eq:HDAF-resolution}
  f(x) \simeq \int \delta_M(x - x';\sigma)f(x')\mathrm{d}x'
\end{equation}
constructed with Hermite polynomials
\begin{equation}
  \label{eq:HDAF}
  \delta_M(x;\sigma)
  = \frac{\exp\left(\frac{-x^2}{2\sigma^2}\right)}{\sqrt{2\pi}\sigma}
  \sum_{m=0}^{M/2}\left(-\frac{1}{4}\right)^m\frac{1}{m!}
  H_{2m}\left(\frac{x}{\sqrt{2}\sigma}\right).
\end{equation}
The identity~\eqref{eq:HDAF-resolution}, allows us to compute the action of any operator $G(\partial_x)$
\begin{equation}
  \label{eq:14}
  G(\partial_x)f(x) \simeq \int G(\partial_x)\delta_M(x - x';\sigma)f(x')\mathrm{d}x'
\end{equation}
and realize that on a uniform grid with $2^n$ points, it can always be written as a linear combination of displacements
\begin{equation}
  G(\partial_x) \to \sum_{i=0}^{2^n-1} (G_{i} D_{+i} + G_{-i}D_{-i}).
\end{equation}

Fortunately, as discussed in Ref.~\cite{gidi2025}, the vectors $\mathbf{G}$ tend to decay rapidly, and this expansion can be truncated to finite order, obtaining the operator as an efficient linear combination of displacements (implemented once more by \rtt{api/function/seemps.register.mpo_weighted_shifts}{mpo\_weighted\_shifts}). In SeeMPS, the function \rtt{api/function/seemps.analysis.hdaf.hdaf_mpo}{hdaf\_mpo} uses these techniques to compute HDAF approximations of derivatives at any order $\partial_x^p$, as well as the kinetic propagator $\exp(-it\partial_x^2/2)$.

\subsection{Function Integration}
\label{sec:integration}
Given a function $f(x)$ encoded as an MPS $\mathbf{v}$, its integral can always be estimated as a weighted sum of the tensor values. Interpreting the quadrature weights as a tensor $\mathbf{w}$, numerical integration reduces to the evaluation of the scalar product between two MPS,
\begin{equation}
  \int_a^b f(x)\mathrm{d}x \simeq \sum_i w_i v_i = \langle \mathbf{w}, \mathbf{v}\rangle
\end{equation}
This formulation naturally extends to multiple dimensions by taking tensor products of one-dimensional quadrature rules (see Sec.~\ref{sec:outer}), allowing multidimensional integrals to be computed efficiently within the MPS framework.

At a high level, SeeMPS provides the routine \rtt{api/function/seemps.analysis.integration.integrate_mps}{\path{integrate_mps}}, which estimates integrals of quantized functions in one or more dimensions. The function supports MPS encodings on equispaced grids, represented by \rtt{api/class/seemps.analysis.mesh.RegularInterval}{\path{RegularInterval}}, and non-equispaced grids based on Chebyshev nodes, represented by \rtt{api/class/seemps.analysis.mesh.ChebyshevInterval}{ChebyshevInterval}.

At a lower level, SeeMPS implements MPS-based quadrature constructions tailored for standard binary quantizations and core arrangements. For equispaced grids, Newton--Cotes rules are provided, including the trapezoidal rule (\rtt{api/function/seemps.analysis.integration.mps_trapezoidal}{mps\_trapezoidal}), Simpson's rule (\rtt{api/function/seemps.analysis.integration.mps_simpson38}{\path{mps_simpson38}}), and higher-order schemes such as the fifth-order rule (\rtt{api/function/seemps.analysis.integration.mps_fifth_order}{\path{mps_fifth_order}}). For Chebyshev grids, spectral quadratures such as the Clenshaw--Curtis (\rtt{api/function/seemps.analysis.integration.mps_clenshaw_curtis}{\path{mps_clenshaw_curtis}}) and Fejér formulas (\rtt{api/function/seemps.analysis.integration.mps_fejer}{\path{mps_fejer}}) are available.

In the general case, quadrature weights can be efficiently constructed using TCI; in SeeMPS, this is supported by the helper function \rtt{api/function/seemps.analysis.integration.quadrature_mesh_to_mps}{\path{quadrature_mesh_to_mps}}, which converts a \rtt{api/class/seemps.analysis.mesh.Mesh}{Mesh} object containing user-defined weights into the corresponding MPS.

\section{Static PDE Solution}
\label{sec:PDE}
SeeMPS can be used to solve both eigenvalue and source problems with Dirichlet zero or periodic boundary conditions. In the first family of problems, we find equations that can be brought into a form such as
\begin{equation}
  \left[D(\partial_x) + V(\mathbf{x})\right]f(\mathbf{x}) = E f(\mathbf{x}),
\end{equation}
with some differential operator $D(\partial_x)$ and some unknown eigenvalue $E$. To address these problems, one needs first to construct an MPO for the equation's left-hand-side operator $H=\left[D(\partial_x) + V(\mathbf{x})\right]$ and use our eigenvalue solvers from Sect.~\ref{sec:eigenvalue}.

In the second family of problems, we find inhomogeneous PDEs that contain a source term, such as
\begin{equation}
  \left[D(\partial_x) + V(\mathbf{x})\right]f(x) = g(\mathbf{x}).
\end{equation}
In this case, we need to encode both the operator $H$ as well as the right-hand-side MPS $g(x) \to \mathbf{x}$ and use the linear solvers from Sect.~\ref{sec:linear-equation}.

Using these methods, we can effectively solve quite many Hermitian and non-Hermitian eigenvalue problems, as well as most linear source problems, provided $f(\mathbf{x})$ has either periodic or zero boundary conditions. The reason for this is that we will be directly reusing the differential operators from Sect.~\ref{sec:differentiation}, which do not contemplate other conditions. We are working to lift this restriction.


\section{Time Evolution Methods}
\label{sec:evolution}
\jg{
By time evolution, we refer to initial value problems where a vector changes according to a linear first order differential equation. A paradigmatic example is the Schrödinger equation for a particle in N dimensions,
\begin{align}
  \partial_t \psi(\mathbf{x},t) = -\mathrm{i}\left[-\frac{1}{2}\nabla^2 + V(\mathbf{x})\right]\psi(\mathbf{x},t),
\end{align}
starting from some initial condition $\psi_0(\mathbf{x})$, but SeeMPS also can treat other time evolution problems associated to many-body physics problems, where the unknown is a many-body wavefunction.}

\jg{All these problems can be generally written as a differential equation
\begin{equation}
  \label{eq:general-linear-evolution}
  \partial_t f(\mathbf{x}, t) = \mathcal{L}\,f(\mathbf{x}, t),
\end{equation}
generated by a linear operator $\mathcal{L}$. Particularly for quantum mechanics, two versions of Hamiltonian evolution are typically used; \textit{real} and \textit{imaginary} time evolution. These correspond to solving~\eqref{eq:general-linear-evolution} for $\mathcal{L} = -\mathrm{i}H$ and $\mathcal{L} = - H$, respectively. However, the same paradigm applies to solving general PDEs, where $\mathcal{L}(\mathbf{x})$ will encode potentials and differential operators in position space.}

\jg{Equations of the form~\eqref{eq:general-linear-evolution} are generally solved by an operator $U(t)$
\begin{equation}
  \mathbf{v}(t) = U(t) \mathbf{v}(0) =\exp(t\mathcal{L}) \mathbf{v}(0).
\end{equation}
Since building $U(t)$ at all times is very costly, there exist many alternative methods for solving these ordinary differential equations that approximate small time steps of the solution in the vector space. SeeMPS generalizes these methods to work with MPS mainly by rewriting the same algorithm using our new MPS-BLAS/LAPACK subroutines, writing the unknowns---functions or many-body wavefunctions---as MPS and the evolution generator $\mathcal{L}$ as MPO. However, as described below, the library also includes other MPS-specific methods.}

\subsection{Runge-Kutta Methods}
\label{sec:explicit}
Runge-Kutta methods approximate the action of the evolution operator with truncated Taylor expansions
\jg{
\begin{equation}
  \mathbf{v}(t+\delta t) = \mathbf{v}(t) + \sum_{p=1}^{P} \frac{1}{p!} \left(\delta t \mathcal{L}\right)^p \mathbf{v}(t)
  + \mathcal{O}(\delta{t}^{p+1}), 
\end{equation}
where $\delta{t}$ denotes the time step and the truncation order $p$ determines both the local error and the computational cost.
}

The SeeMPS library supports four of these explicit methods, with increasing order of precision. They all rely on our BLAS scheme of MPO-MPS and MPS-MPS operations, with implicit \jj{compression} steps controlled by the user. The \rtt{api/function/seemps.evolution.euler}{euler} formula is a  first-order scheme with local truncation error \(\mathcal{O}(\delta{t}^2)\),
\jg{
\begin{equation}
  \mathbf{v}(t+\delta{t}) = \mathbf{v}(t) + \delta{t}\,\mathcal{L} \mathbf{v}(t).
\end{equation}%
}%
The Heun or \rtt{api/function/seemps.evolution.euler2}{euler2} improves this to \(\mathcal{O}(\delta{t}^3)\),
\jg{
\begin{equation}
  \mathbf{v}(t+\delta{t})= \mathbf{v}(t) + \frac{\delta{t}}{2}
  \left[\mathcal{L}\mathbf{v}(t) + \mathcal{L}(\mathbf{v}(t) + \delta{t} \mathcal{L}\mathbf{v}(t))\right].
\end{equation}%
}%
The 4th-order \rtt{api/function/seemps.evolution.runge_kutta}{runge\_kutta} method is the well-known recurrence with $\mathcal{O}(\delta{t}^5)$ error,
\jg{
\begin{align}
  \mathbf{v}(t+\delta{t}) &= \mathbf{v}(t) + \frac{\delta{t}}{6}(\mathbf{k}_1 + 2\mathbf{k}_2 + 2\mathbf{k}_3 + \mathbf{k}_4), \\
  \mathbf{k}_1 &= \mathcal{L}\mathbf{v}(t), \nonumber\\
  \mathbf{k}_2 &= \mathcal{L}\left(\mathbf{v}(t) + \tfrac{\delta{t}}{2} \mathbf{k}_1\right), \nonumber\\
  \mathbf{k}_3 &= \mathcal{L}\left(\mathbf{v}(t) + \tfrac{\delta{t}}{2} \mathbf{k}_2\right), \nonumber\\
  \mathbf{k}_4 &= \mathcal{L}\left(\mathbf{v}(t) + \delta{t} \mathbf{k}_3\right). \nonumber
\end{align}
}
This method, in combination with a 5th order formula, leads to the \rtt{api/function/seemps.evolution.runge_kutta_fehlberg}{runge\_kutta\_felhberg} adaptive step-size solver. This function constructs estimates of the error, adjusting $\delta{t}$ to keep it below a predefined tolerance. Each step may require potential repetitions due to rejected step sizes.

\jj{Two remarks are in order. First, unlike TDVP (Sect.~\ref{sec:TDVP}) and TEBD (Sect.~\ref{sec:TEBD}), \jg{which can preserve norm and unitarity for anti-Hermitian generators $\mathcal{L}=-iH$, Runge-Kutta schemes are dissipative and better suited to general non-Hermitian, dissipative, or PDE-type generators where the problem lacks the structure TEBD requires;} for norm-preserving real-time dynamics of many-body systems we recommend TDVP or TEBD instead.
  Second, the adaptive controller targets only the \textit{discretization} error: decreasing $\delta t$ reduces this error but increases the number of evolution steps and hence the number of compression operations, so the accumulated truncation error tracked by the \texttt{error} field grows. The total error thus has a minimum at an intermediate step size, a trade-off the user can monitor through the reported \texttt{error}.}

\subsection{Implicit Methods}
\label{sec:implicit}
In addition to explicit methods, implicit methods should be considered since they may increase stability for certain applications. SeeMPS includes the Crank-Nicolson method (\rtt{api/function/seemps.evolution.crank_nicolson}{crank\_nicolson}), an $\mathcal{O}(\delta{t}^2)$ algorithm based on the trapezoidal rule that combines the Euler method and its backward version
\jg{
\begin{equation}
    \left(\mathbb{I} - \frac{\delta{t}}{2}\mathcal{L}\right)\mathbf{v}(t+\delta{t}) = \left(\mathbb{I} + \frac{\delta{t}}{2}\mathcal{L}\right)\mathbf{v}(t).
  \end{equation}
}
Each step of the evolution requires us to solve a linear equation, which in this case is done using the CGS method from Sect.~\ref{sec:CGS}.

For stiff problems requiring high-order accuracy, the library provides the \rtt{api/function/seemps.evolution.radau}{radau} integrator, which implements fully implicit Radau IIA methods with $3$ or $5$ stages (order $5$ and $9$, respectively). The algorithm requires us to solve for \jg{an extended} vector of stage derivatives \jg{$\mathbf{K}$} in
\begin{equation}
  \jg{
    \left[\mathbb{I}_s \otimes \mathbb{I} - \delta t (A \otimes \mathcal{L})\right] \jg{\mathbf{K}} = \mathbf{1} \otimes (\mathcal{L} \mathbf{v}(t)),
  }
\end{equation}
where $A$ is the Butcher matrix of size $s\times s$, $s$ is the number of stages, and $\mathbf{1}$ is a vector of ones. This system is solved for $\jg{\mathbf{K}}$, and the state update is obtained by contracting the stage index with the Runge-Kutta weights $\jg{b_j}$,
\begin{equation}
  \mathbf{v}(t+\delta t) = \mathbf{v}(t) + \delta t \sum_{j=1}^s \jg{b_j \mathbf{K}_j}.
\end{equation}

\subsection{\jj{Time-Dependent Variational Principle (TDVP)}}
\label{sec:TDVP}

The time-dependent variational principle (TDVP) provides an optimal approximation to the evolution \jg{generated by $L$} when the state is restricted to the manifold of matrix product states with fixed bond dimension, $\text{MPS}_\chi$. In its original formulation, TDVP was obtained by projecting the Schrödinger equation onto the tangent space of $\text{MPS}_\chi$, resulting in a system of coupled differential equations for the sites of the MPS, which preserve the norm and symmetries of the system, but are numerically stiff to integrate~\cite{haegeman2011}. A later reformulation showed that this projected evolution can be locally integrated, resulting in a stable, sweep-based algorithm closely related to DMRG~\cite{haegeman2016}.

SeeMPS implements the two-site integration scheme, which lifts the fixed-rank limitation of the one-site algorithm, in the \rtt{api/function/seemps.evolution.tdvp}{tdvp} function. The evolution over a time step $\delta t$ is approximated by a Lie-Trotter splitting of the tangent-space projector, resulting in a sequence of local updates. For a bond connecting sites $n$ and $n+1$, the update proceeds in three stages. First, the effective operator $\jg{\mathcal{L}_\text{eff}^{(n, n+1)}}$ is built for the combined two-site tensor $A^{(n, n+1)}$. This tensor is evolved forward in time by a half-step $\delta t/2$,
\begin{align}
    \jg{A^{(n,n+1)}(t + \delta t/2) = \exp\left(\frac{\delta t}{2} \mathcal{L}_{\text{eff}}^{(n,n+1)} \right) A^{(n,n+1)}(t).}
\end{align}
Then, the evolved tensor is decomposed via SVD to restore the canonical MPS form, $A^{(n, n+1)} \to A^{(n)}A^{(n+1)}$, where the singular values are truncated according to a tolerance and maximum bond dimension. To complete the step and advance the center of orthogonality to the next site without overcounting the evolution on the shared subspace, the single-site tensor \jg{$A^{(n+1)}$} is evolved backward in time using the one-site effective operator \jg{$\mathcal{L}_\text{eff}^{(n+1)}$},
\begin{align}
    \jg{A^{(n+1)}(t) = \exp\left(-\frac{\delta t}{2} \mathcal{L}_{\text{eff}}^{(n+1)} \right) A^{(n+1)}(t+\delta t/2).}
\end{align}
The effective \jg{operators $\mathcal{L}_\text{eff}^{(n, n+1)}$ and $\mathcal{L}_\text{eff}^{(n)}$ are} obtained by contracting the MPO with the current left and right environments, managed efficiently by a \texttt{QuadraticForm}. The full-step TDVP evolution is obtained by composing these local solutions in sweeps. In particular, a left-to-right sweep followed by a right-to-left sweep yields a symmetric second-order integrator with local error $\mathcal{O}(\delta t^3)$.

\subsection{\jj{Split-step Method with HDAF Propagators}}
\label{sec:split-step-HDAF}
This integrator is specific to the Schrödinger equation. In particular, when the Hamiltonian can be decomposed into a sum of non-commuting terms, such as $H=-\partial_x^2/2 + V(x)$, the time evolution operator can be efficiently approximated using operator splitting techniques. A standard choice is the Strang splitting,
\begin{align}
    e^{-\jj{\mathrm{i}} \delta t H} = e^{-\jj{\mathrm{i}}\frac{\delta t}{2}V(x) }
    e^{\jj{\mathrm{i}}\frac{\delta t}{2} \partial_x^2}
    e^{-\jj{\mathrm{i}}\frac{\delta t}{2}V(x) } + \mathcal{O}(\delta t^3),
\end{align}
which yields a symplectic integrator equivalent to the Störmer-Verlet scheme. This decomposition separates the evolution into a diagonal potential propagator and a kinetic propagator. SeeMPS implements this scheme in \rtt{api.function.seemps.evolution.hdaf.split\_step}{split\_step}. In this implementation, the diagonal operator $e^{-\jj{\mathrm{i}}\frac{\jg{\delta t}}{2}V(x)}$ is approximated using tensor cross-interpolation techniques (c.f. Sect.~\ref{sec:TCI}). In contrast, the kinetic propagator $e^{\jj{\mathrm{i}}\frac{\jg{\delta t}}{2} \partial_x^2}$ is non-diagonal in the coordinate representation. While standard implementations typically require a transformation to the momentum space using the quantum Fourier transform, SeeMPS adopts an alternative approach based on Hermite Distributed Approximating Functionals~(c.f. Sect.~\ref{sec:HDAF}). This enables the approximation of the kinetic propagator directly in the coordinate basis~\cite{hoffman1991}, where it is represented as a banded MPO constructed as a linear combination of discrete shift operators, with controllable accuracy determined by the truncation order. \jj{The split-step scheme is favored precisely when the Hamiltonian splits as $H=T+V$ into a coordinate-diagonal potential $V(x)$ and a kinetic term $T$ with an efficient propagator, as is typical for Schr\"odinger-type PDEs: it then yields a symplectic, long-time-stable integrator at a fixed cost per step~\cite{gidi2025}.}

\section{Quantum Physics Applications}
\label{sec:quantum}
SeeMPS was originally created to satisfy our research group's needs to study quantum many-body physics with MPS algorithms. Interestingly, quantum applications are mostly covered by the sections we have discussed so far. Indeed, the most relevant uses of MPS include (i) encoding of quantum states and operators as MPS and MPOs, (ii) computation of expected values and properties of quantum states, (iii) calculation of eigenstates of a problem Hamiltonian, and (iv) study of the evolution of a quantum state under a Hamiltonian.

All these tasks are supported in the BLAS, LAPACK, and evolution packages from Sects.~\ref{sec:BLAS}, \ref{sec:LAPACK} and \ref{sec:evolution}. Even the most relevant quantum physics algorithm, the DMRG, is already covered there. Having said this, SeeMPS provides some additional components to help with the study of quantum many-body physics problems, which we now discuss.

\subsection{Long-range Hamiltonian Construction}
\label{sec:Hamiltonian}
\jj{Across SeeMPS, all algorithms---quantum and numerical-analysis alike---are expressed uniformly in terms of matrix product operators.} This allows us to write algorithms that work both for quantum applications and for functional analysis problems. However, while every Hamiltonian can be written in MPO form, the task of cleverly finding this representation for a specific model, such as
\begin{equation}
  H = \sum_i h_i \sigma^z_i + \sum_{ij} J_{ij}\sigma^x_i\sigma^x_j
\end{equation}
can be rather tedious. For this reason, SeeMPS offers a single class, called \rtt{api/class/seemps.hamiltonians.InteractionGraph}{InteractionGraph} that can be used to record all interactions in a model and, in one pass, create the joint MPO.

The class works by collecting the interacting terms in a database. This database is then used to create an artificial MPS---over an auxiliary index that labels the interaction terms---that represents all the interactions and local terms. The MPS is brought into the simplest form possible with a compression stage, and it is then used to recreate the final MPO. \jj{This construction is \textit{exact}: it reproduces the target Hamiltonian to machine precision, and the compression stage merely discovers the minimal bond dimension rather than introducing any approximation.} This technique allows the algorithm to discover that nearest neighbor interactions $\sum_i J_i\sigma^z_i\sigma^z_{i+1}$ can be written with bond dimension two, and \jj{detect other structure such as translation-invariant or finite-range terms that admit a minimal-bond-dimension MPO}. Furthermore, by applying the \jj{compression} stage onto the MPS and not the MPO, we ensure that hermiticity is preserved, even if the final operator is within machine precision of the desired Hamiltonian\footnote{\jj{Working in double precision, the singular value decomposition in LAPACK typically produces} machine-precision errors $\sim10^{-16}$ even for simple problems.}.

\begin{figure}[t!]
\begin{pythoncode}
import numpy as np
from seemps.state import random_uniform_mps
from seemps.operators import MPO
from seemps.hamiltonians import InteractionGraph
from seemps.register import (
    HamiltonianEvolutionLayer,
    ParameterizedLayeredCircuit,
    LocalRotationsLayer,
)
N = 10 # No. qubits
n_layers = 2 # No. layers
sz = np.asarray([[1, 0], [0, -1]])
sx = np.asarray([[0, 1], [1, 0]])
J = np.random.normal(size=(N,N))
H = InteractionGraph([2] * N)
H.add_long_range_interaction(J, sz)
hadamards = MPO.from_local_operators(
    [np.asarray([[1, 1],
                 [1, -1]])/np.sqrt(2)] * N
)
layers = [hadamards]
for _ in range(n_layers):
    layers.append(HamiltonianEvolutionLayer(
        H.to_mpo()))
    layers.append(LocalRotationsLayer(N, sx, same_parameter=True))
U = ParameterizedLayeredCircuit(N, layers)
L = U.parameters_size
assert L == n_layers * (1 + 1)
v = random_uniform_mps(2, N, D=2)
parameters = np.random.normal(size=L)
Uv = U.apply(v, parameters)
\end{pythoncode}
\caption{Pseudocode to \jj{construct and apply} a QAOA\jj{-type parameterized circuit} for an Ising Hamiltonian with random interactions and an example on how to apply it with a selection of parameters. \jj{The pseudocode illustrates only the construction and application of the QAOA circuit. Optimizing the cost function $\langle\mathbf{v}|H_C|\mathbf{v}\rangle$ is then a matter of wrapping these evaluations in the user's preferred optimizer (e.g.\ SciPy's BFGS or Nelder--Mead).}}
\label{fig:qaoa}
\end{figure}

\subsection{\jj{Time-Evolving Block Decimation (TEBD)} and Trotter Methods}
\label{sec:TEBD}
One of the first successful applications of MPS techniques was the simulation of time evolution in 1D many-body systems and open quantum systems. This was made possible by two similar algorithms, the time evolution block decimation algorithm (TEBD)~\cite{vidal2003} and the optimal approximation of Trotterized evolution~\cite{verstraete2004,garcia-ripoll2006}.

In both cases, the starting point is a generator of the real or imaginary time evolution (c.f. Section~\ref{sec:evolution}) that has a nearest-neighbor interaction structure
\begin{equation}
  H = \sum_i H_{i,i+1},
\end{equation}
with a set of local operators $H_{i,i+1}$ that act on neighboring sites, such as $\sigma^x_i \sigma^x_{i+1}+\sigma^y_i\sigma^y_{i+1}$. In this scenario, one may approximate the time evolution with Trotter formulas of varying order, such as
\begin{equation}
  \mathbf{v}(t+\delta{t})
  \simeq U_0(\delta{t}/2) U_1(\delta{t}) U_0(\delta{t}\jj{/2})
\end{equation}
with the even and odd operators
\begin{equation}
  U_{k}(\epsilon)=\exp\left(-\jj{\mathrm{i}}\epsilon \sum_i H_{2i+k,2i+k+1}\right).
\end{equation}

Obviously, this is one of the few algorithms that cannot operate with an MPO description. For this reason, SeeMPS offers some classes that describe nearest-neighbor interacting terms, such as \rtt{api/class/seemps.hamiltonians.ConstantNNHamiltonian}{ConstantNNHamiltonian} and \rtt{api/class/seemps.hamiltonians.ConstantTIHamiltonian}{ConstantTIHamiltonian}. These specific Hamiltonians can be fed into the 2nd and 3rd order Trotter expansion objects \rtt{api/class/seemps.evolution.Trotter2ndOrder}{Trotter2ndOrder}and \rtt{api/class/seemps.evolution.Trotter3rdOrder}{Trotter3rdOrder}, to evolve an MPS for a brief period of time.

\subsection{Quantum Computer Emulation}
\label{sec:quantum-computer}
MPS can be used to classically simulate weakly entangled computations, as already pointed out by Vidal in Ref.~\cite{vidal2003}. SeeMPS's architecture is particularly well suited for this task: (i) a single \texttt{MPO} can represent a layer of quantum gates that are encoded with low bond dimension; (ii) several of these layers can be chained into an \texttt{MPOList} to represent a quantum circuit and (iii) our standard contraction and \jj{compression} routines can be used to study how circuits act on an \texttt{MPS} quantum register with moderate resources in a \jj{convenient and flexible} way.

With these tools, SeeMPS implements a framework for variational quantum algorithms that is quite flexible. The main class is the \rtt{api/class/seemps.register.ParameterizedLayeredCircuit}{ParameterizedLayeredCircuit}, which is a collection of unitary operations with and without parameters. Among the latter, we find layers with local gates on each qubit (\rtt{api/class/seemps.register.LocalRotationsLayer}{LocalRotationsLayer}), layers with entangling gates in a 1D nearest-neighbourgh architecture (\rtt{api/class/seemps.register.TwoQubitGatesLayer}{TwoQubitGatesLayer}), a multi-qubit gate implementing evolution with a generic MPO Hamiltonian (\rtt{api/class/seemps.register.HamiltonianEvolutionLayer}{HamiltonianEvolutionLayer}), and generic, parameter free layers encoded as MPO (\rtt{api/class/seemps.register.ParameterFreeMPO}{ParameterFreeMPO}).

Together, these layers may be combined to create complex parameterized circuits, such as a hardware-efficient 1D VQE ansatz (\rtt{api/class/seemps.register.VQECircuit}{VQECircuit}), or a generic QAOA variational circuit for Ising models (\rtt{api/class/seemps.register.IsingQAOACircuit}{IsingQAOACircuit}). As an illustrative example, Figure~\ref{fig:qaoa} offers the pseudocode to construct a QAOA parameterized circuit from a random Ising model without magnetic field.

With these tools, it is possible to implement rather sophisticated investigations. \jj{In particular, the study of variational states produced by QAOA circuits reported in Refs.~\cite{diez-valle2023, diez-valle2025} made use of SeeMPS; the same machinery also supports variational quantum eigensolvers (see Refs.~\cite{bharti2022, tilly2022} for the general method) and quantum machine-learning models.} Still to be written, however, is a compatibility layer that allows encoding more generic circuits, such as those coming from Qiskit or Pennylane, into the SeeMPS library formalism.

\section{Library Architecture and Usage}
\label{sec:architecture}
\begin{figure}[t]
  \centering
  \includegraphics[width=\linewidth]{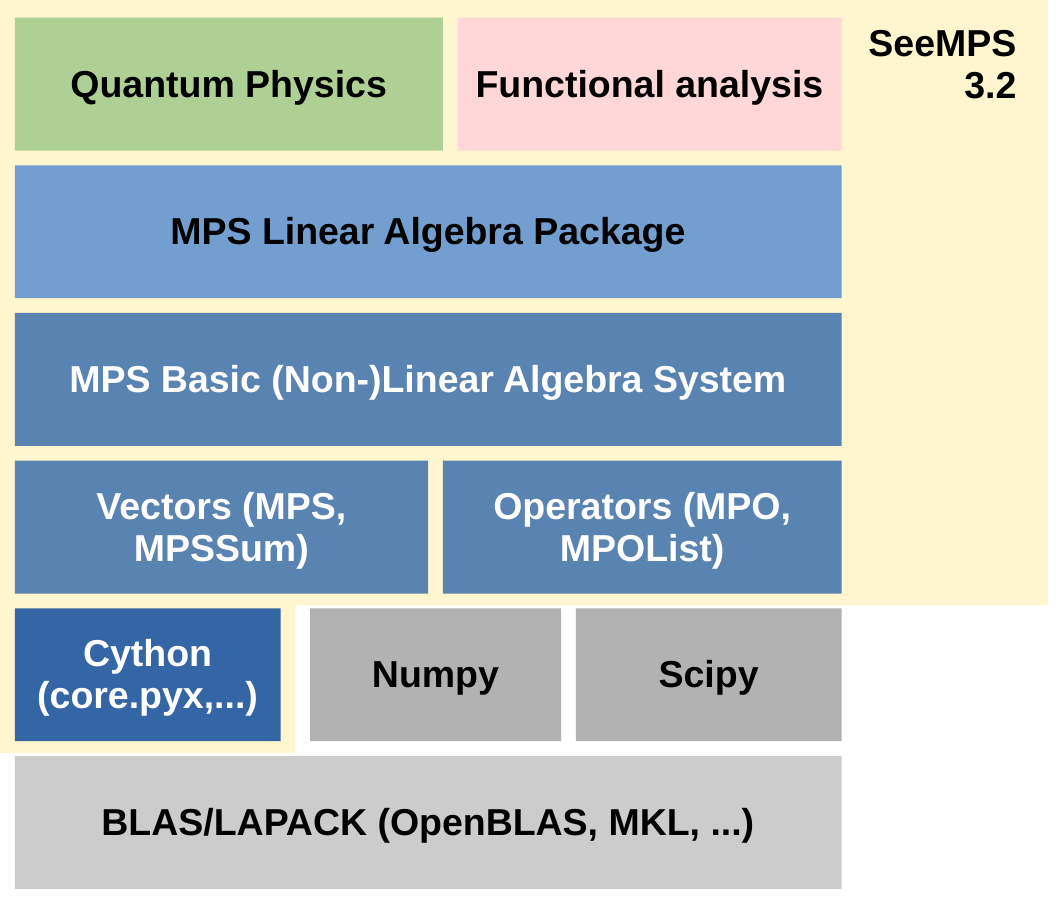}
  \caption{\jj{Layered architecture of SeeMPS. A low-level BLAS/LAPACK core (OpenBLAS or Intel MKL, provided through NumPy and SciPy) together with a few high-efficiency Cython components underpins the MPS and MPO data structures. On top of these sit the MPS-BLAS operations (Sect.~\ref{sec:BLAS}), the intermediate-level MPS-LAPACK algorithms (Sect.~\ref{sec:LAPACK}), and, at the highest level, the applied-mathematics and quantum-physics applications.}}
  \label{fig:architecture}
\end{figure}

The SeeMPS library follows a layered architecture, described in Figure~\ref{fig:architecture}. The core of the library is formed by a traditional BLAS/LAPACK component---typically OpenBLAS (in PyPI) or Intel's MKL (in Conda or Conda-Forge)---that is provided by NumPy and SciPy, and consumed also by a few high-efficiency components developed in Cython 3.0 or later~\cite{cython2025}. On top of these low-level components, we construct the basic data structures for the MPS and MPO objects, followed by the BLAS that operates on them (Sect.~\ref{sec:BLAS}). It then follows the intermediate-level algorithms (Sect.~\ref{sec:LAPACK}) and then the applications to operate with applied mathematics and quantum physics problems.

As shown in Table~\ref{tab:lines-of-code}, SeeMPS takes around 18000 lines of code, separated in about 10600 for the library itself and 7500 for unit testing. Of the library, 8\% is made of Cython code that provides optimized functions to (i) decompose a tensor into two tensors using different truncation strategies, (ii) implement key multidimensional tensor contractions used in various algorithms, (iii) compute an MPS's canonical form, and (iv) implement a scalar product among MPS.

The choice of which components to optimize is based on a careful study of performance bottlenecks in real-world applications, both in quantum physics and numerical analysis. This led to rewriting functions that were coded in pure Python ---e.g., the routine that searches for the optimal split of two tensors based on error tolerances---to replace them with tightly optimized C-like loops. By translating the code to C with the Cython compiler, the final functions are significantly faster and consume less memory, making the library useful for actual applications. While this library is still less performant than other (and leaner) versions of the MPS architecture implemented in C++ by the authors, it is still competitive with similar implementations in Julia and Python. Furthermore, we are not aware of any performant library that provides such a comprehensive coverage of MPS/TT algorithms for numerical analysis applications.

The development of SeeMPS places strong emphasis on static analysis and systematic testing. Approximately 43\% of the codebase is devoted to unit tests, which collectively exercise about 94\% of the production code. These tests prevent regressions in critical components—such as the MPS compression algorithm—and enforce the invariants associated with the code---e.g., a canonical form MPS is always formed by isometries, algorithms satisfy predefined error bounds, existence of run-time error checks, exceptions and warnings, etc. These tests have also been useful to detect regressions in core components, such as errors and instabilities in the SVD routine from Intel MKL.

In addition, the library makes extensive use of type annotations. While Python is a dynamically typed language, consistent use of type hints provides self-documenting interfaces and enables early detection of mismatches in argument passing and return values, thereby improving reliability and long-term maintainability. These type annotations are checked using both MyPy~\cite{mypy2025} and basedpyright~\cite{basedpyright2025}, and the code is linted using ruff~\cite{ruff2025}.

All these checks (static analysis, unit tests, linting, etc.) are enforced with the help of GitHub's CI/CD workflows, under every commit and every pull request, to ensure consistency of the code and its contributions. \jj{Because SeeMPS deliberately supports a broad range of NumPy and SciPy versions across Linux, Windows and macOS---rather than pinning to a narrow release window, which would frequently conflict with the platform-specific builds that users already have installed---the full test suite is periodically re-run on GitHub and on our development machines against current NumPy and SciPy releases, so that any incompatibility is detected early.}

\begin{table}[t]
\centering
\begin{tabular}{lrr}
\hline
Module & Python (LOC) & Cython (LOC) \\
\hline
Core & 564 & 812 \\
BLAS & 2743 & 0 \\
LAPACK & 983 & 0 \\
Loading & 2883 & 0 \\
Differentiation & 263 & 0 \\
Integration & 391 & 0 \\
Interpolation & 380 & 0 \\
Evolution & 712 & 0 \\
Quantum & 885 & 0 \\
Unit tests & 7503 & 0 \\
\hline
\textbf{Total} & 17307 & 812 \\
\hline
\end{tabular}
\caption{\jj{Lines of code by library section. Unit tests exercise $\sim$94\% of statements.}}
\label{tab:lines-of-code}
\end{table}
As of version \jg{3.2}, which is the version documented in this article, SeeMPS is available in Python's package index PyPI, and can be directly installed using
\begin{verbatim}
    pip install seemps
\end{verbatim}
and it can be listed with this name as a dependency in \verb|pyproject.toml| files. Precompiled binaries are regularly built in GitHub for \jj{x86-64 Linux and Windows, as well as for Apple-silicon (arm64) macOS}. The library is distributed under a \jj{permissive} MIT license and has been used in various works~\cite{garcia-ripoll2021, garcia-molina2024, gidi2025, rodriguez-aldavero2025} from our group, as well as in real-world quantum science applications~\cite{shi2018, feiguin2020}.

The SeeMPS project is developed in \href{https://github.com/juanjosegarciaripoll/seemps2}{GitHub}, a platform that provides us with CI/CD for testing and easy mechanisms for \href{https://github.com/juanjosegarciaripoll/seemps2/issues}{\texttt{bug reporting}} and contributions. The library's documentation is built on every minor release and uploaded to \href{https://seemps.readthedocs.io}{ReadTheDocs}. All functions and classes linked in this document can be found in those pages. The folder \href{https://github.com/juanjosegarciaripoll/seemps2/tree/main/examples}{\texttt{examples}} in the library contains Jupyter notebooks that illustrate the practical usage of all these techniques, both in quantum and quantum-inspired applications.

\section{Examples and Benchmarking}
\label{sec:examples-and-benchmarking}

\jg{In this section we solve some paradigmatic problems with the library and use them as evidence of its efficiency. Sections~\ref{sec:example-dmrg}--\ref{sec:example-pde-solution} address the ground state of a quantum spin chain, its time evolution, and the solution of a diffusion equation, spanning both quantum many-body physics and quantum-inspired numerical analysis. Section~\ref{sec:benchmarking} then reuses these examples to benchmark the runtime and accuracy of the library against the ITensor framework.}

\begin{figure}[h!]
\begin{pythoncode}
import numpy as np
from seemps.hamiltonians import ConstantNNHamiltonian
from seemps.optimization import dmrg

n = 10  # No. spins
g = 1.0  # Transverse field

sz = np.asarray([[1, 0], [0, -1]]) / 2
sx = np.asarray([[0, 1], [1, 0]]) / 2
sz_sz = np.kron(sz, sz)

# Build TFIM Hamiltonian
H = ConstantNNHamiltonian(n, dimension=2)
for i in range(n - 1):
    H.add_interaction_term(i, -sz_sz)
for i in range(n):
    H.add_local_term(i, -g * sx)
Hmpo = H.to_mpo()
    
# Find ground state
result = dmrg(Hmpo)
E0, psi0 = result.energy, result.state
\end{pythoncode}
\caption{\jg{Pseudocode to construct the transverse-field Ising Hamiltonian as an MPO and compute its ground-state energy $E_0$ and state $\psi_0$ via DMRG.}}
\label{fig:dmrg-tfim}
\end{figure}

\subsection{DMRG}
\label{sec:example-dmrg}

\jg{Our first example is a standard quantum many-body task: computing the ground state of a local Hamiltonian with the most representative MPS algorithm, DMRG (Sect.~\ref{sec:DMRG}). We consider the transverse-field Ising model (TFIM),
\begin{align*}
  H = -\sum_{\langle i,j\rangle}\sigma_i^z\sigma_j^z - \sum_j\sigma_j^x,
\end{align*}
a quantum spin chain in which the sum over nearest neighbors $\langle i,j\rangle$ competes with a transverse field.}

\jg{Figure~\ref{fig:dmrg-tfim} builds the Hamiltonian as an MPO from its local and two-body terms---deliberately avoiding the translationally invariant Hamiltonian classes, so the code generalizes readily to non-uniform terms---and hands it to the \texttt{dmrg} function to search for the lowest-energy state.}

\subsection{TDVP}
\label{sec:example-tdvp}
\begin{figure}[t!]
\begin{pythoncode}
import numpy as np
from seemps.state import product_state
from seemps.evolution import tdvp

T = 1.0     # Evolution time
steps = 50  # No. of steps

# Hmpo is the TFIM Hamiltonian MPO
local_state = np.array([1.0, 1.0]) / np.sqrt(2.0)
psi_t0 = product_state(local_state, length=Hmpo.size)
psi_T = tdvp(-1j * Hmpo, (0.0, T), psi_t0, steps=steps)
\end{pythoncode}
\caption{\jg{Pseudocode to evolve a product state under the transverse-field Ising Hamiltonian using $2-$site TDVP.}}
\label{fig:tdvp-tfim}
\end{figure}
\jg{This second example reuses the TFIM Hamiltonian, now evolving an unentangled product state in time with the time-dependent variational principle (Sect.~\ref{sec:TDVP}). Assuming the MPO Hamiltonian from before, the code passes it together with the MPS initial state to \rtt{api/function/seemps.evolution.tdvp}{tdvp}, supplying the evolution generator explicitly as $-\mathrm{i}H$ following our earlier conventions.}

\subsection{PDE Solution}
\label{sec:example-pde-solution}
\begin{figure}[t!]
\begin{pythoncode}
import numpy as np
from seemps.analysis.mesh import QuantizedInterval, Mesh, mps_to_mesh_matrix
from seemps.analysis.cross import BlackBoxLoadMPS, cross_maxvol
from seemps.analysis.derivatives import finite_differences_mpo
from seemps.evolution import runge_kutta

a, b = -0.5, 0.5 # Space interval
N = 10      # Grid points = 2^N
steps = 50  # No. RK4 steps
nu = 0.01   # Diffusion coefficient

dx = (b - a) / 2 ** N
dt = 0.5 * dx ** 2 / nu # diffusion stability condition
T = steps * dt

# Load initial MPS without dense vectors
interval = QuantizedInterval(a, b, N)
mesh = Mesh([interval])
M = mps_to_mesh_matrix([N])

def f_t0(x):
    return np.exp(-0.5 * (x[0] / 0.1) ** 2)
bb = BlackBoxLoadMPS(f_t0, mesh, M, [2] * N)
psi_t0 = cross_maxvol(bb).mps

# Evolve
D2 = finite_differences_mpo(2, interval)
psi_T = runge_kutta(nu * D2, (0.0, T), psi_t0, steps=steps)
\end{pythoncode}
\caption{\jg{Pseudocode to solve the diffusion PDE~\eqref{eq:diffusion-pde} on a quantized grid, loading a Gaussian initial state via tensor cross-interpolation (TCI) and evolving it with a finite-difference Laplacian MPO and the fourth-order Runge-Kutta.}}
\label{fig:rk4-diffusion}
\end{figure}
\jg{The diffusion (heat) equation,
\begin{align}
  \label{eq:diffusion-pde}
  \frac{\partial u}{\partial t} - \nu\frac{\partial^2 u}{\partial x^2} = 0,
\end{align}
is a prototypical linear PDE, which we solve with a standard fourth-order Runge-Kutta scheme (Sect.~\ref{sec:explicit}). Unlike the previous cases, it is not the evolution of a quantum system but a purely numerical-analysis task.}

\jg{In Figure~\ref{fig:rk4-diffusion} the code defines the spatial grid, loads a Gaussian initial condition through tensor cross-interpolation (Sect.~\ref{sec:TCI}) to avoid instantiating a large dense vector, and builds the generator---a non-Hermitian finite-difference operator---which, together with the initial state, is handed to the solver \rtt{api/function/seemps.evolution.runge_kutta}{runge\_kutta}. The time step is fixed for stability; this is intrinsic to the diffusion PDE with explicit integrators, not a limitation of the library or the MPS formalism.}

\subsection{Benchmarking}
\label{sec:benchmarking}
\jg{We reuse these three examples as a benchmark of the accuracy and efficiency of SeeMPS across its application domains. Alongside the standard library we test a variant in which the small Cython core is replaced by a pure-Python implementation, to gauge the kernel's value, and we compare against the widely adopted Julia implementation of ITensor.}

\jg{This comparison must be taken with a grain of salt, as the two libraries serve different purposes: ITensor is a general framework for tensor-network computations, supporting techniques such as exact quantum-number conservation and sparse tensors that SeeMPS does not provide, whereas SeeMPS targets numerical-analysis applications (with some quantum-algorithm use), where those specializations matter less. Concretely, we benchmark against the ITensorMPS module; because of this different focus, the comparison required custom differential operators and PDE solvers on the ITensor side, written to parallel their SeeMPS counterparts as closely as possible. The benchmarking scripts can be found in SeeMPS's examples folder.}

\begin{figure*}[htb!]
  \centering
  \includegraphics[width=\textwidth]{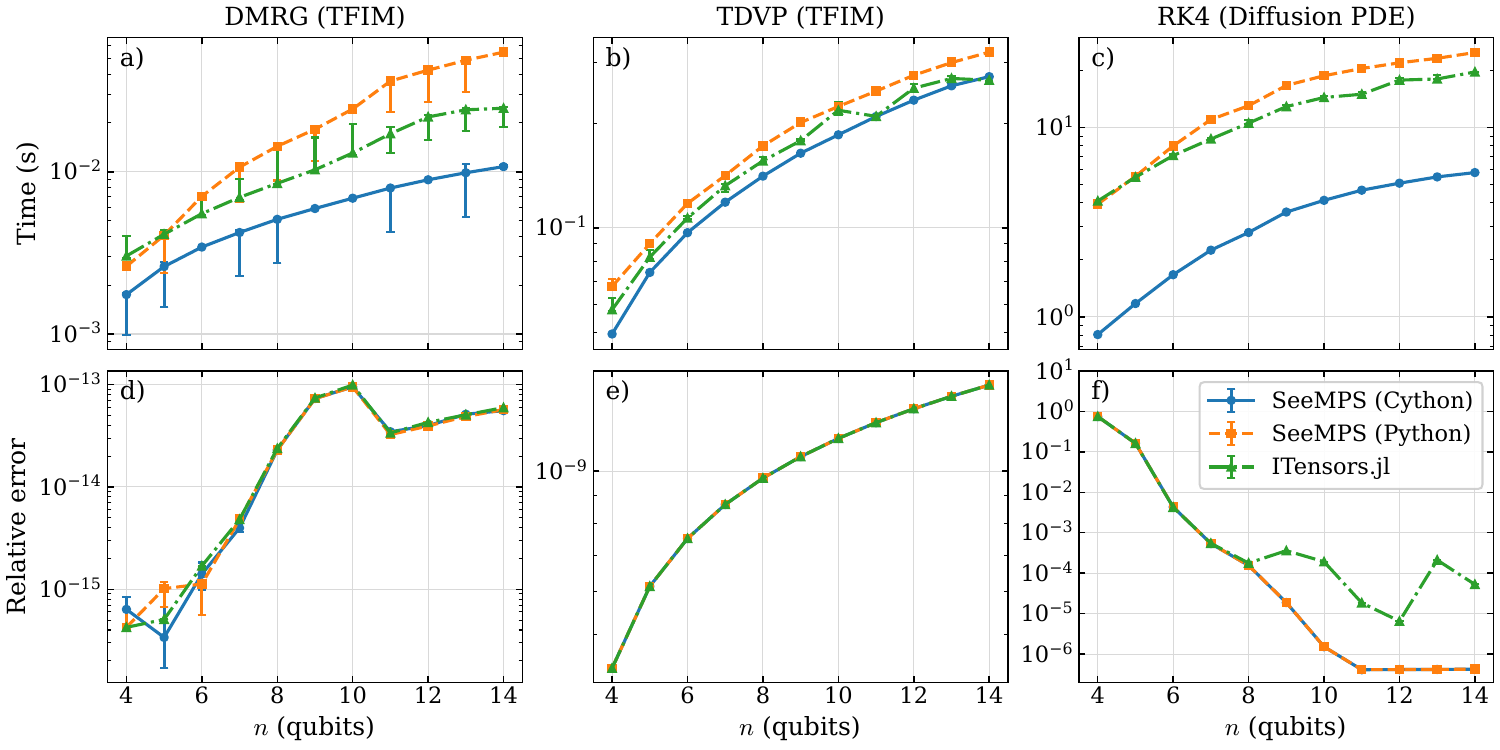}
  \caption{\jg{Comparison of SeeMPS (Python and Cython cores) against the ITensor library, in terms of runtime (top) and relative error (bottom), for three use cases: Performing a ground-state search for a transverse-field Ising model (TFIM) Hamiltonian (a) and (d), time evolution under the same Hamiltonian (b) and (e), and time evolution for a Diffusion PDE (c) and (f).\label{fig:benchmark}}}
\end{figure*}

\jg{The benchmark results are summarized on Figure~\ref{fig:benchmark} for the $3$ examples presented before in this section. All benchmarks share the same relative truncation tolerance of the singular values, $\varepsilon^2 / (n - 1)$ with $\varepsilon=10^{-6}$, and do not restrict the maximum tensor dimensions. However, we should note that each library defaults to a different compression mechanism: while SeeMPS uses variational compression, ITensor relies on SVD truncation. Each task was run once just before benchmarking, to rule out JIT compilation or any other initialization source of slowness. We report the median and interquartile range of the runtimes and accuracies from each use case, after $101$ repetitions.}

\jg{For the DMRG ground-state search [Fig.~\ref{fig:benchmark}(a,d)] we triggered early exit on convergence with a \texttt{DMRGObserver} in ITensor. Runtimes are comparable, with the Cython core leading and ITensor sharing the same scaling in the number of qubits; the Python core scales differently at small sizes---dominated by language overhead---before catching up. The errors are indistinguishable across libraries and backends, and well within the requested tolerance.}

\jg{For the TDVP time evolution [Fig.~\ref{fig:benchmark}(b,e)] the runtime is dominated by the matrix exponential of the local one- and two-site Hamiltonians, giving similar runtimes and identical errors for the three libraries; on the ITensor side we used the \texttt{applyexp} updater, faster than the default here.}

\jg{The diffusion-PDE integration [Fig.~\ref{fig:benchmark}(c,f)], with $\nu = 0.01$ and $1000$ time steps of size $\Delta t = (\Delta x)^2/20\nu$, stresses the combined cost of the integrator's inner routines: MPO--MPS application, MPS addition, and compression. Here the differences are clearest: the Python core and ITensor yield similar runtimes, while the Cython core keeps the same scaling with a four-to-five-fold reduction in runtime. Both SeeMPS backends reach the same accuracy---set by the Runge--Kutta time-discretization error at low qubit counts, then floored at the requested truncation tolerance---whereas ITensor, at the same tolerance, shows an accuracy degradation that grows above the integration error for $9$ qubits or more, consistent with truncation-error accumulation. We are not aware of any asymmetry between the Runge--Kutta implementations used for the two libraries, besides the default compression algorithms.}

\jg{In summary, the Cython backend is justified on efficiency grounds, and SeeMPS is competitive in runtime---at equal accuracy---with ITensor, a well-established performant library in the field.}

\section{Conclusion}
\label{sec:conclusion}
SeeMPS is a simple yet powerful library for implementing quantum and quantum-inspired algorithms based on MPS and MPO tensor networks. It offers a low-level complete package for linear-algebra programming using MPS/MPO. On top of this package, the library provides an extensive collection of algorithms, enabling applications in (i) the study and emulation quantum many-body physics, (ii) the emulation of quantum computations, and (iii) providing a complete framework for numerical analysis.

While many tensor-network frameworks exist with competitive performance and advanced features---e.g., sparse-tensor and symmetry handling---most are designed around specific application domains. Generic platforms include ITensor~\cite{fishman2022}, a widely adopted Julia framework offering a rich ecosystem for tensor-network algorithms, and emerging platforms such as Tenet~\cite{tenet_jl} and other infrastructure-level libraries~\cite{lyakh2022exatn, roberts2019tensornetwork}. A second group focuses primarily on quantum many-body computations, including Python libraries such as Quimb~\cite{gray2018quimb} and TenPy~\cite{hauschild2018efficient}, which provides highly optimized implementations of DMRG and TEBD, alongside packages such as MPSKit.jl~\cite{mpskit_jl}. Finally, libraries oriented toward numerical analysis include the TT-Toolbox~\cite{tt_toolbox} MATLAB package, an early reference implementation for TT methods; Teneva~\cite{chertkov2023black}, providing Python implementations using a functional style; and TensorKrowch~\cite{monturiol2024tensorkrowch}, targeting the QTT representation and manipulation of machine-learning models. A broader overview of the software landscape can be found in the literature~\cite{sehlstedt2025software}.

In contrast, SeeMPS is conceived as an architecturally simple Python library explicitly focused on \jj{the MPS/TT formalism and its quantized (QTT) form}, built with an algorithm-centric design. Despite its simplicity, it achieves competitive performance through an optimized Cython core. Moreover, its BLAS/LAPACK abstraction provides a complete framework for numerical analysis that is quite accessible for users with a linear algebra background, such as undergraduate students in physics, mathematics, or engineering. As shown throughout this work, conventional algorithms that operate on matrices and vectors can be translated almost verbatim from Python/NumPy or MATLAB into this programming environment. For these reasons, SeeMPS is well suited to advance the field of tensor-train and quantized tensor-train numerical methods---a domain that presently lacks similar unifying tools.

\paragraph{Acknowledgements\jj{~--}} 

This work has been supported by the Spanish Ministry of Science, Innovation and Universities and the State Research Agency (MICIU\slash AEI\slash 10.13039\slash 501100011033) through Projects No. PID2021-127968NB-I00, No. PDC2022-133486-I00, and No. PCI2025-163266, co-funded by the European Union ``NextGenerationEU''\slash PRTR. It has also been supported by the Ministry for Digital Transformation and of Civil Service of the Spanish Government through the QUANTUM ENIA project call - Quantum Spain project, and by the European Union through the Recovery, Transformation and Resilience Plan - NextGenerationEU within the framework of the Digital Spain 2026 Agenda. Furthermore, this work has received funding from the EuroHPC Joint Undertaking under project QEC4QEA (Ref. 101194322). JJGR and JJRA acknowledge support from the CSIC Interdisciplinary Thematic Platform (PTI) Quantum Technologies (PTI-QTEP+). PGM acknowledges support by projects QuCv (AIA2025-163435-C43) and SUKIDI (PID2024-157778OB-I00), and grant QML-CV SDC007\slash 25\slash 000001 from the Catalan Department of Enterprise and Labor.




\bibliographystyle{elsarticle-num}
\bibliography{bibliography, bibliography_software}

@article{adak2025,
  title = {Tensor {{Network Space-Time Spectral Collocation Method}} for {{Solving}} the {{Nonlinear Convection Diffusion Equation}}},
  author = {Adak, Dibyendu and Danis, M. Engin and Truong, Duc P. and Rasmussen, Kim {\O}. and Alexandrov, Boian S.},
  year = 2025,
  month = mar,
  journal = {J Sci Comput},
  volume = {103},
  number = {2},
  pages = {46},
  issn = {1573-7691},
  doi = {10.1007/s10915-025-02860-x},
  urldate = {2026-01-11},
  langid = {english}
}

@article{Barratt2022,
  title = {Improvements to Gradient Descent Methods for Quantum Tensor Network Machine Learning},
  author = {Barratt, Fergus and Dborin, James and Wright, Lewis},
  year = 2022,
  journal = {arXiv e-prints},
  volume = {arXiv:2203.03366},
  publisher = {arXiv},
  doi = {10.48550/ARXIV.2203.03366},
  copyright = {Creative Commons Attribution 4.0 International}
}

@misc{basedpyright2025,
  title = {Basedpyright: Static Type Checker},
  howpublished = {\url{https://docs.basedpyright.com/latest/}},
  year = {2025}
}

@article{bharti2022,
  title = {Noisy Intermediate-Scale Quantum Algorithms},
  author = {Bharti, Kishor and {Cervera-Lierta}, Alba and Kyaw, Thi Ha and Haug, Tobias and {Alperin-Lea}, Sumner and Anand, Abhinav and Degroote, Matthias and Heimonen, Hermanni and Kottmann, Jakob S. and Menke, Tim and Mok, Wai-Keong and Sim, Sukin and Kwek, Leong-Chuan and {Aspuru-Guzik}, Al{\'a}n},
  year = 2022,
  month = feb,
  journal = {Rev. Mod. Phys.},
  volume = {94},
  number = {1},
  pages = {015004},
  publisher = {American Physical Society},
  doi = {10.1103/RevModPhys.94.015004},
  urldate = {2026-01-10}
}

@article{bigoni2016,
  title = {Spectral {{Tensor-Train Decomposition}}},
  author = {Bigoni, Daniele and {Engsig-Karup}, Allan P. and Marzouk, Youssef M.},
  year = 2016,
  month = jan,
  journal = {SIAM J. Sci. Comput.},
  volume = {38},
  number = {4},
  pages = {A2405-A2439},
  issn = {1064-8275, 1095-7197},
  doi = {10.1137/15M1036919},
  urldate = {2026-01-11},
  langid = {english}
}

@article{chen2023,
  title = {Quantum {{Fourier Transform Has Small Entanglement}}},
  author = {Chen, Jielun and Stoudenmire, E.M. and White, Steven R.},
  year = 2023,
  month = oct,
  journal = {PRX Quantum},
  volume = {4},
  number = {4},
  pages = {040318},
  publisher = {American Physical Society},
  doi = {10.1103/PRXQuantum.4.040318},
  urldate = {2025-12-26}
}

@article{chen2026,
  title = {Direct Interpolative Construction of the Discrete {{Fourier}} Transform as a Matrix Product Operator},
  author = {Chen, Jielun and Lindsey, Michael},
  year = 2026,
  month = feb,
  journal = {Applied and Computational Harmonic Analysis},
  volume = {81},
  pages = {101817},
  issn = {10635203},
  doi = {10.1016/j.acha.2025.101817},
  urldate = {2026-01-11},
  langid = {english}
}

@article{cirac2021,
  title = {Matrix Product States and Projected Entangled Pair States: {{Concepts}}, Symmetries, Theorems},
  shorttitle = {Matrix Product States and Projected Entangled Pair States},
  author = {Cirac, J. Ignacio and {P{\'e}rez-Garc{\'i}a}, David and Schuch, Norbert and Verstraete, Frank},
  year = 2021,
  month = dec,
  journal = {Rev. Mod. Phys.},
  volume = {93},
  number = {4},
  pages = {045003},
  publisher = {American Physical Society},
  doi = {10.1103/RevModPhys.93.045003},
  urldate = {2025-12-16}
}

@article{clenshaw1955,
  title = {A Note on the Summation of {{Chebyshev}} Series},
  author = {Clenshaw, C. W.},
  year = 1955,
  journal = {Math. Comp.},
  volume = {9},
  number = {51},
  pages = {118--120},
  issn = {0025-5718, 1088-6842},
  doi = {10.1090/S0025-5718-1955-0071856-0},
  urldate = {2026-01-11},
  langid = {english}
}

@misc{coppersmith2002,
  title = {An Approximate {{Fourier}} Transform Useful in Quantum Factoring},
  author = {Coppersmith, D.},
  year = 2002,
  month = jan,
  number = {arXiv:quant-ph/0201067},
  eprint = {quant-ph/0201067},
  publisher = {arXiv},
  doi = {10.48550/arXiv.quant-ph/0201067},
  urldate = {2025-12-26},
  archiveprefix = {arXiv}
}

@misc{cython2025,
  title = {Cython},
  howpublished = {\url{https://cython.org/}},
  year = {2025}
}

@article{diez-valle2023,
  title = {Quantum {{Approximate Optimization Algorithm Pseudo-Boltzmann States}}},
  author = {{D{\'i}ez-Valle}, Pablo and Porras, Diego and {Garc{\'i}a-Ripoll}, Juan Jos{\'e}},
  year = 2023,
  month = feb,
  journal = {Phys. Rev. Lett.},
  volume = {130},
  number = {5},
  pages = {050601},
  publisher = {American Physical Society},
  doi = {10.1103/PhysRevLett.130.050601},
  urldate = {2026-01-10}
}

@misc{diez-valle2025,
  title = {Universal {{Resources}} for {{QAOA}} and {{Quantum Annealing}}},
  author = {{D{\'i}ez-Valle}, Pablo and {G{\'o}mez-Ruiz}, Fernando J. and Porras, Diego and {Garc{\'i}a-Ripoll}, Juan Jos{\'e}},
  year = 2025,
  month = jun,
  number = {arXiv:2506.03241},
  eprint = {2506.03241},
  primaryclass = {quant-ph},
  publisher = {arXiv},
  doi = {10.48550/arXiv.2506.03241},
  urldate = {2026-01-10},
  archiveprefix = {arXiv}
}

@article{dolgov2020,
  title = {Parallel Cross Interpolation for High-Precision Calculation of High-Dimensional Integrals},
  author = {Dolgov, Sergey and Savostyanov, Dmitry},
  year = 2020,
  month = jan,
  journal = {Computer Physics Communications},
  volume = {246},
  pages = {106869},
  issn = {00104655},
  doi = {10.1016/j.cpc.2019.106869},
  urldate = {2026-01-11},
  langid = {english}
}

@book{driscoll2014,
  title = {Chebfun Guide},
  author = {Driscoll, Tobin and Hale, Nicholas and Trefethen, Lloyd N.},
  year = 2014,
  publisher = {Pafnuty Publications},
  address = {Oxford}
}

@article{dukelsky1998,
  title = {Equivalence of the Variational Matrix Product Method and the Density Matrix Renormalization Group Applied to Spin Chains},
  author = {Dukelsky, J. and {Mart{\'i}n-Delgado}, M. A. and Nishino, T. and Sierra, G.},
  year = 1998,
  month = aug,
  journal = {EPL},
  volume = {43},
  number = {4},
  pages = {457},
  publisher = {IOP Publishing},
  issn = {0295-5075},
  doi = {10.1209/epl/i1998-00381-x},
  urldate = {2025-12-27},
  langid = {english}
}

@article{feiguin2020,
  title = {Qubit-Photon Corner States in All Dimensions},
  author = {Feiguin, Adrian and {Garc{\'i}a-Ripoll}, Juan Jos{\'e} and {Gonz{\'a}lez-Tudela}, Alejandro},
  year = 2020,
  month = apr,
  journal = {Phys. Rev. Res.},
  volume = {2},
  number = {2},
  pages = {023082},
  publisher = {American Physical Society},
  doi = {10.1103/PhysRevResearch.2.023082},
  urldate = {2025-12-29}
}

@article{fishman2022,
  title = {The {{ITensor Software Library}} for {{Tensor Network Calculations}}},
  author = {Fishman, Matthew and White, Steven and Stoudenmire, Edwin Miles},
  year = 2022,
  month = aug,
  journal = {SciPost Physics Codebases},
  pages = {004},
  issn = {2949-804X},
  doi = {10.21468/SciPostPhysCodeb.4},
  urldate = {2025-12-07},
  langid = {english}
}

@article{garcia-molina2022,
  title = {Quantum {{Fourier}} Analysis for Multivariate Functions and Applications to a Class of {{Schr\"odinger-type}} Partial Differential Equations},
  author = {{Garc{\'i}a-Molina}, Paula and {Rodr{\'i}guez-Mediavilla}, Javier and {Garc{\'i}a-Ripoll}, Juan Jos{\'e}},
  year = 2022,
  month = jan,
  journal = {Phys. Rev. A},
  volume = {105},
  number = {1},
  pages = {012433},
  issn = {2469-9926, 2469-9934},
  doi = {10.1103/PhysRevA.105.012433},
  urldate = {2025-12-26},
  langid = {english}
}

@misc{garcia-molina2024,
  title = {Global Optimization of {{MPS}} in Quantum-Inspired Numerical Analysis},
  author = {{Garc{\'i}a-Molina}, Paula and Tagliacozzo, Luca and {Garc{\'i}a-Ripoll}, Juan Jos{\'e}},
  year = 2024,
  month = may,
  number = {arXiv:2303.09430},
  eprint = {2303.09430},
  primaryclass = {quant-ph},
  publisher = {arXiv},
  doi = {10.48550/arXiv.2303.09430},
  urldate = {2025-12-26},
  archiveprefix = {arXiv}
}

@phdthesis{garcia-molina2025,
  title = {Quantum Computing and Quantum-Inspired Numerical Methods. {{Application}} to Problems in Condensed Matter Physics and Other Fields},
  author = {{Garc{\'i}a-Molina}, Paula},
  year = 2025,
  school = {Universidad Aut{\'o}noma de Madrid},
  type = {Ph.{{D}}. thesis},
  url = {http://hdl.handle.net/10486/719805},
  urldate = {2026-07-08}
}

@article{garcia-ripoll2006,
  title = {Time Evolution of {{Matrix Product States}}},
  author = {{Garc{\'i}a-Ripoll}, Juan Jos{\'e}},
  year = 2006,
  month = dec,
  journal = {New J. Phys.},
  volume = {8},
  number = {12},
  pages = {305},
  issn = {1367-2630},
  doi = {10.1088/1367-2630/8/12/305},
  urldate = {2025-12-26},
  langid = {english}
}

@article{garcia-ripoll2021,
  title = {Quantum-Inspired Algorithms for Multivariate Analysis: From Interpolation to Partial Differential Equations},
  shorttitle = {Quantum-Inspired Algorithms for Multivariate Analysis},
  author = {{Garc{\'i}a-Ripoll}, Juan Jos{\'e}},
  year = 2021,
  month = apr,
  journal = {Quantum},
  volume = {5},
  pages = {431},
  publisher = {Verein zur F\"orderung des Open Access Publizierens in den Quantenwissenschaften},
  doi = {10.22331/q-2021-04-15-431},
  urldate = {2025-12-26},
  langid = {british}
}

@article{gidi2025,
  title = {Pseudospectral Method for Solving {{PDEs}} Using Matrix Product States},
  author = {Gidi, Jorge and {Garc{\'i}a-Molina}, Paula and Tagliacozzo, Luca and {Garc{\'i}a-Ripoll}, Juan Jos{\'e}},
  year = 2025,
  month = oct,
  journal = {Journal of Computational Physics},
  volume = {539},
  pages = {114228},
  issn = {0021-9991},
  doi = {10.1016/j.jcp.2025.114228},
  urldate = {2025-12-26}
}

@article{goreinov1997,
  title = {A Theory of Pseudoskeleton Approximations},
  author = {Goreinov, S.A. and Tyrtyshnikov, E.E. and Zamarashkin, N.L.},
  year = 1997,
  month = aug,
  journal = {Linear Algebra and its Applications},
  volume = {261},
  number = {1-3},
  pages = {1--21},
  issn = {00243795},
  doi = {10.1016/S0024-3795(96)00301-1},
  urldate = {2026-01-11},
  copyright = {https://www.elsevier.com/tdm/userlicense/1.0/},
  langid = {english}
}

@inbook{goreinov2010,
  title = {How to {{Find}} a {{Good Submatrix}}},
  booktitle = {Matrix {{Methods}}: {{Theory}}, {{Algorithms}} and {{Applications}}},
  author = {Goreinov, S. A. and Oseledets, I. V. and Savostyanov, D. V. and Tyrtyshnikov, E. E. and Zamarashkin, N. L.},
  year = 2010,
  month = apr,
  pages = {247--256},
  publisher = {WORLD SCIENTIFIC},
  doi = {10.1142/9789812836021_0015},
  urldate = {2026-01-11},
  collaborator = {Olshevsky, Vadim and Tyrtyshnikov, Eugene},
  isbn = {978-981-283-601-4 978-981-283-602-1},
  langid = {english}
}

@article{Gorodetsky2018b,
  title = {Gradient-Based Optimization for Regression in the Functional Tensor-Train Format},
  author = {Gorodetsky, Alex A. and Jakeman, John D.},
  year = 2018,
  journal = {Journal of Computational Physics},
  volume = {374},
  pages = {1219--1238},
  issn = {0021-9991},
  doi = {10.1016/j.jcp.2018.08.010}
}

@article{gourianov2022,
  title = {A Quantum-Inspired Approach to Exploit Turbulence Structures},
  author = {Gourianov, Nikita and Lubasch, Michael and Dolgov, Sergey and {van den Berg}, Quincy Y. and Babaee, Hessam and Givi, Peyman and Kiffner, Martin and Jaksch, Dieter},
  year = 2022,
  month = jan,
  journal = {Nat Comput Sci},
  volume = {2},
  number = {1},
  pages = {30--37},
  publisher = {Nature Publishing Group},
  issn = {2662-8457},
  doi = {10.1038/s43588-021-00181-1},
  urldate = {2025-12-28},
  copyright = {2022 The Author(s), under exclusive licence to Springer Nature America, Inc.},
  langid = {english}
}

@misc{grover2002,
  title = {Creating Superpositions That Correspond to Efficiently Integrable Probability Distributions},
  author = {Grover, Lov and Rudolph, Terry},
  year = 2002,
  month = aug,
  number = {arXiv:quant-ph/0208112},
  eprint = {quant-ph/0208112},
  publisher = {arXiv},
  doi = {10.48550/arXiv.quant-ph/0208112},
  urldate = {2026-01-11},
  archiveprefix = {arXiv}
}

@article{haegeman2011,
  title = {Time-{{Dependent Variational Principle}} for {{Quantum Lattices}}},
  author = {Haegeman, Jutho and Cirac, J. Ignacio and Osborne, Tobias J. and Pi{\v z}orn, Iztok and Verschelde, Henri and Verstraete, Frank},
  year = 2011,
  month = aug,
  journal = {Phys. Rev. Lett.},
  volume = {107},
  number = {7},
  pages = {070601},
  publisher = {American Physical Society},
  doi = {10.1103/PhysRevLett.107.070601},
  urldate = {2025-12-30}
}

@article{haegeman2016,
  title = {Unifying Time Evolution and Optimization with Matrix Product States},
  author = {Haegeman, Jutho and Lubich, Christian and Oseledets, Ivan and Vandereycken, Bart and Verstraete, Frank},
  year = 2016,
  month = oct,
  journal = {Phys. Rev. B},
  volume = {94},
  number = {16},
  pages = {165116},
  publisher = {American Physical Society},
  doi = {10.1103/PhysRevB.94.165116},
  urldate = {2025-12-30}
}

@article{halimeh2015,
  title = {Chebyshev Matrix Product State Approach for Time Evolution},
  author = {Halimeh, Jad C. and Kolley, Fabian and McCulloch, Ian P.},
  year = 2015,
  month = sep,
  journal = {Phys. Rev. B},
  volume = {92},
  number = {11},
  pages = {115130},
  issn = {1098-0121, 1550-235X},
  doi = {10.1103/PhysRevB.92.115130},
  urldate = {2026-01-11},
  copyright = {http://link.aps.org/licenses/aps-default-license},
  langid = {english}
}

@article{hoffman1991,
  title = {Analytic Banded Approximation for the Discretized Free Propagator},
  author = {Hoffman, David K. and Nayar, Naresh and Sharafeddin, Omar A. and Kouri, D. J.},
  year = 1991,
  month = oct,
  journal = {J. Phys. Chem.},
  volume = {95},
  number = {21},
  pages = {8299--8305},
  publisher = {American Chemical Society},
  issn = {0022-3654},
  doi = {10.1021/j100174a052},
  urldate = {2025-12-28}
}

@article{holzner2011,
  title = {Chebyshev Matrix Product State Approach for Spectral Functions},
  author = {Holzner, Andreas and Weichselbaum, Andreas and McCulloch, Ian P. and Schollw{\"o}ck, Ulrich and Von Delft, Jan},
  year = 2011,
  month = may,
  journal = {Phys. Rev. B},
  volume = {83},
  number = {19},
  pages = {195115},
  issn = {1098-0121, 1550-235X},
  doi = {10.1103/PhysRevB.83.195115},
  urldate = {2026-01-11},
  copyright = {http://link.aps.org/licenses/aps-default-license},
  langid = {english}
}

@article{hur2023,
  title = {Generative Modeling via Tensor Train Sketching},
  author = {Hur, YoonHaeng and Hoskins, Jeremy G. and Lindsey, Michael and Stoudenmire, E.M. and Khoo, Yuehaw},
  year = 2023,
  month = nov,
  journal = {Applied and Computational Harmonic Analysis},
  volume = {67},
  pages = {101575},
  issn = {10635203},
  doi = {10.1016/j.acha.2023.101575},
  urldate = {2026-01-11},
  langid = {english}
}

@article{iblisdir2007,
  title = {Matrix Product States Algorithms and Continuous Systems},
  author = {Iblisdir, S. and Or{\'u}s, R. and Latorre, J. I.},
  year = 2007,
  month = mar,
  journal = {Phys. Rev. B},
  volume = {75},
  number = {10},
  pages = {104305},
  publisher = {American Physical Society},
  doi = {10.1103/PhysRevB.75.104305},
  urldate = {2026-01-11}
}

@article{jobst2024,
  title = {Efficient {{MPS}} Representations and Quantum Circuits from the {{Fourier}} Modes of Classical Image Data},
  author = {Jobst, Bernhard and Shen, Kevin and Riofr{\'i}o, Carlos A. and Shishenina, Elvira and Pollmann, Frank},
  year = 2024,
  month = dec,
  journal = {Quantum},
  volume = {8},
  pages = {1544},
  publisher = {Verein zur F\"orderung des Open Access Publizierens in den Quantenwissenschaften},
  doi = {10.22331/q-2024-12-03-1544},
  urldate = {2025-12-26},
  langid = {british}
}

@article{khoromskij2010,
  title = {Quantics-{{TT Collocation Approximation}} of {{Parameter-Dependent}} and {{Stochastic Elliptic PDEs}}},
  author = {Khoromskij, B. N. and Oseledets, I.},
  year = 2010,
  month = jan,
  journal = {Computational Methods in Applied Mathematics},
  volume = {10},
  number = {4},
  pages = {376--394},
  publisher = {De Gruyter},
  issn = {1609-9389},
  doi = {10.2478/cmam-2010-0023},
  urldate = {2025-12-28},
  copyright = {De Gruyter expressly reserves the right to use all content for commercial text and data mining within the meaning of Section 44b of the German Copyright Act.},
  langid = {english}
}

@article{khoromskij2011,
  title = {{{QTT}} Approximation of Elliptic Solution Operators in Higher Dimensions},
  author = {Khoromskij, B. N. and Oseledets, I. V.},
  year = 2011,
  month = jun,
  volume = {26},
  number = {3},
  pages = {303--322},
  publisher = {De Gruyter},
  issn = {1569-3988},
  doi = {10.1515/rjnamm.2011.017},
  urldate = {2025-12-28},
  chapter = {Russian Journal of Numerical Analysis and Mathematical Modelling},
  copyright = {De Gruyter expressly reserves the right to use all content for commercial text and data mining within the meaning of Section 44b of the German Copyright Act.},
  langid = {english}
}

@article{khoromskij2011a,
  title = {O(Dlog\,{{N}})-{{Quantics Approximation}} of {{N-d Tensors}} in {{High-Dimensional Numerical Modeling}}},
  author = {Khoromskij, Boris N.},
  year = 2011,
  month = oct,
  journal = {Constr Approx},
  volume = {34},
  number = {2},
  pages = {257--280},
  issn = {1432-0940},
  doi = {10.1007/s00365-011-9131-1},
  urldate = {2026-01-11},
  langid = {english}
}

@article{kuhner1999,
  title = {Dynamical Correlation Functions Using the Density Matrix Renormalization Group},
  author = {K{\"u}hner, Till D. and White, Steven R.},
  year = 1999,
  month = jul,
  journal = {Phys. Rev. B},
  volume = {60},
  number = {1},
  pages = {335--343},
  publisher = {American Physical Society},
  doi = {10.1103/PhysRevB.60.335},
  urldate = {2025-12-27}
}

@misc{latorre2005,
  title = {Image Compression and Entanglement},
  author = {Latorre, Jose I.},
  year = 2005,
  month = oct,
  number = {arXiv:quant-ph/0510031},
  eprint = {quant-ph/0510031},
  publisher = {arXiv},
  doi = {10.48550/arXiv.quant-ph/0510031},
  urldate = {2026-01-11},
  archiveprefix = {arXiv}
}

@misc{lindsey2024,
  title = {Multiscale Interpolative Construction of Quantized Tensor Trains},
  author = {Lindsey, Michael},
  year = 2024,
  month = apr,
  number = {arXiv:2311.12554},
  eprint = {2311.12554},
  publisher = {arXiv},
  doi = {10.48550/arXiv.2311.12554},
  urldate = {2026-01-11},
  archiveprefix = {arXiv}
}

@article{lubasch2018,
  title = {Multigrid Renormalization},
  author = {Lubasch, Michael and Moinier, Pierre and Jaksch, Dieter},
  year = 2018,
  month = nov,
  journal = {Journal of Computational Physics},
  volume = {372},
  pages = {587--602},
  issn = {0021-9991},
  doi = {10.1016/j.jcp.2018.06.065},
  urldate = {2025-12-28}
}

@article{mikhalev2018,
  title = {Rectangular Maximum-Volume Submatrices and Their Applications},
  author = {Mikhalev, A. and Oseledets, I.V.},
  year = 2018,
  month = feb,
  journal = {Linear Algebra and its Applications},
  volume = {538},
  pages = {187--211},
  issn = {00243795},
  doi = {10.1016/j.laa.2017.10.014},
  urldate = {2026-01-11},
  langid = {english}
}

@article{Monro1979,
  title = {Interpolation by Fast Fourier and Chebyshev Transforms},
  author = {Monro, Donlad M.},
  year = 1979,
  journal = {International Journal for Numerical Methods in Engineering},
  volume = {14},
  number = {11},
  eprint = {https://onlinelibrary.wiley.com/doi/pdf/10.1002/nme.1620141109},
  pages = {1679--1692},
  doi = {10.1002/nme.1620141109}
}

@misc{mypy2025,
  title = {MyPy},
  howpublished = {\url{https://mypy-lang.org/}},
  year = {2025}
}

@article{nunez-fernandez2022,
  title = {Learning {{Feynman Diagrams}} with {{Tensor Trains}}},
  author = {N{\'u}{\~n}ez Fern{\'a}ndez, Yuriel and Jeannin, Matthieu and Dumitrescu, Philipp T. and Kloss, Thomas and Kaye, Jason and Parcollet, Olivier and Waintal, Xavier},
  year = 2022,
  month = nov,
  journal = {Phys. Rev. X},
  volume = {12},
  number = {4},
  pages = {041018},
  issn = {2160-3308},
  doi = {10.1103/PhysRevX.12.041018},
  urldate = {2026-01-11},
  langid = {english}
}

@article{nunez-fernandez2025,
  title = {Learning Tensor Networks with Tensor Cross Interpolation: {{New}} Algorithms and Libraries},
  shorttitle = {Learning Tensor Networks with Tensor Cross Interpolation},
  author = {N{\'u}{\~n}ez Fern{\'a}ndez, Yuriel and Ritter, Marc K and Jeannin, Matthieu and Li, Jheng-Wei and Kloss, Thomas and Louvet, Thibaud and Terasaki, Satoshi and Parcollet, Olivier and Von Delft, Jan and Shinaoka, Hiroshi and Waintal, Xavier},
  year = 2025,
  month = mar,
  journal = {SciPost Phys.},
  volume = {18},
  number = {3},
  pages = {104},
  issn = {2542-4653},
  doi = {10.21468/SciPostPhys.18.3.104},
  urldate = {2026-01-11}
}

@article{Nyquist1928,
  title = {Certain {{Topics}} in {{Telegraph Transmission Theory}}},
  author = {Nyquist, Harry},
  year = 1928,
  journal = {Transactions of the American Institute of Electrical Engineers},
  volume = {47},
  number = {2},
  pages = {617--644},
  doi = {10.1109/t-aiee.1928.5055024}
}

@article{oseledets2010,
  title = {Approximation of \$2\textasciicircum d\textbackslash times2\textasciicircum d\$ {{Matrices Using Tensor Decomposition}}},
  author = {Oseledets, I. V.},
  year = 2010,
  month = jun,
  journal = {SIAM Journal on Matrix Analysis and Applications},
  publisher = {{Society for Industrial and Applied Mathematics}},
  doi = {10.1137/090757861},
  urldate = {2025-12-28},
  copyright = {Copyright \copyright{} 2010 Society for Industrial and Applied Mathematics},
  langid = {english}
}

@article{oseledets2010a,
  title = {{{TT-cross}} Approximation for Multidimensional Arrays},
  author = {Oseledets, Ivan and Tyrtyshnikov, Eugene},
  year = 2010,
  month = jan,
  journal = {Linear Algebra and its Applications},
  volume = {432},
  number = {1},
  pages = {70--88},
  issn = {00243795},
  doi = {10.1016/j.laa.2009.07.024},
  urldate = {2026-01-11},
  copyright = {https://www.elsevier.com/tdm/userlicense/1.0/},
  langid = {english}
}

@article{oseledets2011,
  title = {Tensor-{{Train Decomposition}}},
  author = {Oseledets, I. V.},
  year = 2011,
  month = sep,
  journal = {SIAM Journal on Scientific Computing},
  publisher = {{Society for Industrial and Applied Mathematics}},
  doi = {10.1137/090752286},
  urldate = {2026-01-01},
  copyright = {Copyright \copyright{} 2011 Society for Industrial and Applied Mathematics},
  langid = {english}
}

@article{oseledets2013,
  title = {Constructive {{Representation}} of {{Functions}} in {{Low-Rank Tensor Formats}}},
  author = {Oseledets, I. V.},
  year = 2013,
  month = feb,
  journal = {Constr Approx},
  volume = {37},
  number = {1},
  pages = {1--18},
  issn = {1432-0940},
  doi = {10.1007/s00365-012-9175-x},
  urldate = {2026-01-11},
  langid = {english}
}

@article{ostlund1995,
  title = {Thermodynamic {{Limit}} of {{Density Matrix Renormalization}}},
  author = {{\"O}stlund, Stellan and Rommer, Stefan},
  year = 1995,
  month = nov,
  journal = {Phys. Rev. Lett.},
  volume = {75},
  number = {19},
  pages = {3537--3540},
  publisher = {American Physical Society},
  doi = {10.1103/PhysRevLett.75.3537},
  urldate = {2025-12-27}
}

@article{pareja-monturiol2025,
  title = {Tensorization of Neural Networks for Improved Privacy and Interpretability},
  author = {Pareja Monturiol, Jos{\'e} Ram{\'o}n and {Pozas-Kerstjens}, Alejandro and {P{\'e}rez-Garc{\'i}a}, David},
  year = 2025,
  month = dec,
  journal = {SciPost Physics Core},
  volume = {8},
  number = {4},
  pages = {095},
  issn = {2666-9366},
  doi = {10.21468/SciPostPhysCore.8.4.095},
  urldate = {2026-01-15},
  langid = {english}
}

@article{Pollock2021,
  title = {Extrapolating the Arnoldi Algorithm to Improve Eigenvector Convergence},
  author = {Pollock, Sara and Scott, L.Ridgway},
  year = 2021,
  journal = {International Journal of Numerical Analysis and Modeling},
  volume = {18},
  number = {5},
  pages = {712--721},
  issn = {2617-8710}
}

@article{porras2006,
  title = {Renormalization Algorithm for the Calculation of Spectra of Interacting Quantum Systems},
  author = {Porras, D. and Verstraete, F. and Cirac, J. I.},
  year = 2006,
  month = jan,
  journal = {Phys. Rev. B},
  volume = {73},
  number = {1},
  pages = {014410},
  publisher = {American Physical Society},
  doi = {10.1103/PhysRevB.73.014410},
  urldate = {2026-01-01}
}

@article{ritter2024,
  title = {Quantics {{Tensor Cross Interpolation}} for {{High-Resolution Parsimonious Representations}} of {{Multivariate Functions}}},
  author = {Ritter, Marc K. and N{\'u}{\~n}ez Fern{\'a}ndez, Yuriel and Wallerberger, Markus and Von Delft, Jan and Shinaoka, Hiroshi and Waintal, Xavier},
  year = 2024,
  month = jan,
  journal = {Phys. Rev. Lett.},
  volume = {132},
  number = {5},
  pages = {056501},
  issn = {0031-9007, 1079-7114},
  doi = {10.1103/PhysRevLett.132.056501},
  urldate = {2026-01-11},
  langid = {english}
}

@misc{rodriguez-aldavero2025,
  title = {Chebyshev Approximation and Composition of Functions in Matrix Product States for Quantum-Inspired Numerical Analysis},
  author = {{Rodr{\'i}guez-Aldavero}, Juan Jos{\'e} and {Garc{\'i}a-Molina}, Paula and Tagliacozzo, Luca and {Garc{\'i}a-Ripoll}, Juan Jos{\'e}},
  year = 2025,
  month = feb,
  number = {arXiv:2407.09609},
  eprint = {2407.09609},
  primaryclass = {quant-ph},
  publisher = {arXiv},
  doi = {10.48550/arXiv.2407.09609},
  urldate = {2025-12-28},
  archiveprefix = {arXiv}
}

@misc{ruff2025,
  title = {Ruff: An Extremely Fast Python Linter and Code Formatter},
  howpublished = {\url{https://docs.astral.sh/ruff}},
  year = {2025}
}

@misc{ryzhakov2022,
  title = {Constructive {{TT-representation}} of the Tensors given as Index Interaction Functions with Applications},
  author = {Ryzhakov, Gleb and Oseledets, Ivan},
  year = 2022,
  month = jun,
  number = {arXiv:2206.03832},
  eprint = {2206.03832},
  publisher = {arXiv},
  doi = {10.48550/arXiv.2206.03832},
  urldate = {2026-01-11},
  archiveprefix = {arXiv}
}

@inproceedings{savostyanov2011,
  title = {Fast Adaptive Interpolation of Multi-Dimensional Arrays in Tensor Train Format},
  booktitle = {The 2011 {{International Workshop}} on {{Multidimensional}} ({{nD}}) {{Systems}}},
  author = {Savostyanov, Dmitry and Oseledets, Ivan},
  year = 2011,
  month = sep,
  pages = {1--8},
  doi = {10.1109/nDS.2011.6076873},
  urldate = {2026-01-15}
}

@article{schollwock2005,
  title = {The Density-Matrix Renormalization Group},
  author = {Schollw{\"o}ck, U.},
  year = 2005,
  month = apr,
  journal = {Rev. Mod. Phys.},
  volume = {77},
  number = {1},
  pages = {259--315},
  publisher = {American Physical Society},
  doi = {10.1103/RevModPhys.77.259},
  urldate = {2025-12-27}
}

@article{schollwock2011,
  title = {The Density-Matrix Renormalization Group in the Age of Matrix Product States},
  author = {Schollw{\"o}ck, Ulrich},
  year = 2011,
  month = jan,
  journal = {Annals of Physics},
  series = {January 2011 {{Special Issue}}},
  volume = {326},
  number = {1},
  pages = {96--192},
  issn = {0003-4916},
  doi = {10.1016/j.aop.2010.09.012},
  urldate = {2025-12-27}
}

@article{Shannon1949,
  title = {Communication in the {{Presence}} of {{Noise}}},
  author = {Shannon, Claude E.},
  year = 1949,
  journal = {Proceedings of the IRE},
  volume = {37},
  number = {1},
  pages = {10--21},
  doi = {10.1109/jrproc.1949.232969}
}

@article{shi2018,
  title = {Ultrastrong {{Coupling Few-Photon Scattering Theory}}},
  author = {Shi, Tao and Chang, Yue and {Garc{\'i}a-Ripoll}, Juan Jos{\'e}},
  year = 2018,
  month = apr,
  journal = {Phys. Rev. Lett.},
  volume = {120},
  number = {15},
  pages = {153602},
  publisher = {American Physical Society},
  doi = {10.1103/PhysRevLett.120.153602},
  urldate = {2025-12-29}
}

@misc{snrd,
  title = {Smooth Noise-Robust Differentiators},
  author = {Holoborodko, Pavel},
  year = 2008
}

@article{sozykin2022,
  title = {{{TTOpt}}: {{A Maximum Volume Quantized Tensor Train-based Optimization}} and Its {{Application}} to {{Reinforcement Learning}}},
  shorttitle = {{{TTOpt}}},
  author = {Sozykin, Konstantin and Chertkov, Andrei and Schutski, Roman and Phan, Anh-Huy and Cichocki, Andrzej S. and Oseledets, Ivan},
  year = 2022,
  month = dec,
  journal = {Advances in Neural Information Processing Systems},
  volume = {35},
  pages = {26052--26065},
  urldate = {2026-01-11},
  langid = {english}
}

@article{tilly2022,
  title = {The {{Variational Quantum Eigensolver}}: {{A}} Review of Methods and Best Practices},
  shorttitle = {The {{Variational Quantum Eigensolver}}},
  author = {Tilly, Jules and Chen, Hongxiang and Cao, Shuxiang and Picozzi, Dario and Setia, Kanav and Li, Ying and Grant, Edward and Wossnig, Leonard and Rungger, Ivan and Booth, George H. and Tennyson, Jonathan},
  year = 2022,
  month = nov,
  journal = {Physics Reports},
  series = {The {{Variational Quantum Eigensolver}}: A Review of Methods and Best Practices},
  volume = {986},
  pages = {1--128},
  issn = {0370-1573},
  doi = {10.1016/j.physrep.2022.08.003},
  urldate = {2026-01-10}
}

@article{verstraete2004,
  title = {Matrix {{Product Density Operators}}: {{Simulation}} of {{Finite-Temperature}} and {{Dissipative Systems}}},
  shorttitle = {Matrix {{Product Density Operators}}},
  author = {Verstraete, F. and {Garc{\'i}a-Ripoll}, J. J. and Cirac, J. I.},
  year = 2004,
  month = nov,
  journal = {Phys. Rev. Lett.},
  volume = {93},
  number = {20},
  pages = {207204},
  publisher = {American Physical Society},
  doi = {10.1103/PhysRevLett.93.207204},
  urldate = {2025-12-30}
}

@article{verstraete2004a,
  title = {Density {{Matrix Renormalization Group}} and {{Periodic Boundary Conditions}}: {{A Quantum Information Perspective}}},
  shorttitle = {Density {{Matrix Renormalization Group}} and {{Periodic Boundary Conditions}}},
  author = {Verstraete, F. and Porras, D. and Cirac, J. I.},
  year = 2004,
  month = nov,
  journal = {Phys. Rev. Lett.},
  volume = {93},
  number = {22},
  pages = {227205},
  publisher = {American Physical Society},
  doi = {10.1103/PhysRevLett.93.227205},
  urldate = {2026-01-01}
}

@article{vidal2003,
  title = {Efficient {{Classical Simulation}} of {{Slightly Entangled Quantum Computations}}},
  author = {Vidal, Guifr{\'e}},
  year = 2003,
  month = oct,
  journal = {Phys. Rev. Lett.},
  volume = {91},
  number = {14},
  pages = {147902},
  publisher = {American Physical Society},
  doi = {10.1103/PhysRevLett.91.147902},
  urldate = {2025-12-16}
}

@article{vidal2004,
  title = {Efficient {{Simulation}} of {{One-Dimensional Quantum Many-Body Systems}}},
  author = {Vidal, Guifr{\'e}},
  year = 2004,
  month = jul,
  journal = {Phys. Rev. Lett.},
  volume = {93},
  number = {4},
  pages = {040502},
  publisher = {American Physical Society},
  doi = {10.1103/PhysRevLett.93.040502},
  urldate = {2026-01-01}
}

@article{Wang2020,
  title = {Anomaly Detection with Tensor Networks},
  author = {Wang, Jinhui and Roberts, Chase and Vidal, Guifr{\'e} and Leichenauer, Stefan},
  year = 2020,
  journal = {arXiv e-prints},
  volume = {arXiv:2006.02516},
  publisher = {arXiv},
  doi = {10.48550/ARXIV.2006.02516},
  copyright = {arXiv.org perpetual, non-exclusive license}
}

@article{white1992,
  title = {Density Matrix Formulation for Quantum Renormalization Groups},
  author = {White, Steven R.},
  year = 1992,
  month = nov,
  journal = {Phys. Rev. Lett.},
  volume = {69},
  number = {19},
  pages = {2863--2866},
  publisher = {American Physical Society},
  doi = {10.1103/PhysRevLett.69.2863},
  urldate = {2025-12-27}
}

@article{white1993,
  title = {Density-Matrix Algorithms for Quantum Renormalization Groups},
  author = {White, Steven R.},
  year = 1993,
  month = oct,
  journal = {Phys. Rev. B},
  volume = {48},
  number = {14},
  pages = {10345--10356},
  publisher = {American Physical Society},
  doi = {10.1103/PhysRevB.48.10345},
  urldate = {2025-12-27}
}

@article{wiki:bicgs2025,
  title = {Biconjugate Gradient Method},
  year = 2025,
  month = oct,
  journal = {Wikipedia},
  urldate = {2025-12-27},
  copyright = {Creative Commons Attribution-ShareAlike License},
  langid = {english}
}

@article{wiki:cgs2025,
  title = {Conjugate Gradient Method},
  year = 2025,
  month = dec,
  journal = {Wikipedia},
  urldate = {2025-12-27},
  copyright = {Creative Commons Attribution-ShareAlike License},
  langid = {english}
}

@article{zalka1998,
  title = {Simulating Quantum Systems on a Quantum Computer},
  author = {Zalka, Christof},
  year = 1998,
  month = jan,
  journal = {Proc. A},
  volume = {454},
  number = {1969},
  pages = {313--322},
  issn = {1364-5021},
  doi = {10.1098/rspa.1998.0162},
  urldate = {2025-12-28}
}

@article{ramasesha1997,
title = "Low-lying electronic excitations and nonlinear optic properties of polymers via symmetrized density matrix renormalization group method",
author = "S. Ramasesha and Pati, \{Swapan K.\} and Krishnamurthy, \{H. R.\} and Z. Shuai and Br{\'e}das, \{J. L.\}",
year = "1997",
month = mar,
day = "15",
doi = "10.1016/s0379-6779(97)80136-1",
language = "English (US)",
volume = "85",
pages = "1019--1022",
journal = "Synthetic Metals",
issn = "0379-6779",
publisher = "Elsevier B.V.",
number = "1-3",
}

@misc{tenet_jl,
  title        = {Tenet.jl: A Julia Library for Tensor Networks},
  author       = {{Barcelona Supercomputing Center}},
  howpublished = {\url{https://bsc-quantic.github.io/Tenet.jl}},
}

@article{lyakh2022exatn,
  title={Exatn: Scalable gpu-accelerated high-performance processing of general tensor networks at exascale},
  author={Lyakh, Dmitry I and Nguyen, Thien and Claudino, Daniel and Dumitrescu, Eugene and McCaskey, Alexander J},
  journal={Frontiers in Applied Mathematics and Statistics},
  volume={8},
  pages={838601},
  year={2022},
  publisher={Frontiers Media SA}
}

@misc{roberts2019tensornetwork,
      title={TensorNetwork: A Library for Physics and Machine Learning}, 
      author={Chase Roberts and Ashley Milsted and Martin Ganahl and Adam Zalcman and Bruce Fontaine and Yijian Zou and Jack Hidary and Guifre Vidal and Stefan Leichenauer},
      year={2019},
      eprint={1905.01330},
      archivePrefix={arXiv},
      primaryClass={physics.comp-ph}
}

@article{gray2018quimb,
    title={quimb: a python library for quantum information and many-body calculations},
    author={Gray, Johnnie},
    journal={Journal of Open Source Software},
    year = {2018},
    volume={3}, number={29}, pages={819},
    doi={10.21105/joss.00819},
}

@article{hauschild2018efficient,
  title={Efficient numerical simulations with tensor networks: Tensor Network Python (TeNPy)},
  author={Hauschild, Johannes and Pollmann, Frank},
  journal={SciPost Physics Lecture Notes},
  pages={005},
  year={2018}
}

@misc{mpskit_jl,
  author       = {Van Damme, M. and Devos, L. and Haegeman, J.},
  title        = {MPSKit.jl},
  year         = {2025},
  doi          = {10.5281/zenodo.10654901},
  howpublished = {Zenodo}
}

@misc{tt_toolbox,
  author       = {Oseledets, Ivan},
  title        = {TT-Toolbox},
  year         = {2026},
  howpublished = {\url{https://github.com/oseledets/TT-Toolbox}},
  note         = {GitHub repository}
}

@article{chertkov2023black,
  title={Black box approximation in the tensor train format initialized by ANOVA decomposition},
  author={Chertkov, Andrei and Ryzhakov, Gleb and Oseledets, Ivan},
  journal={SIAM Journal on Scientific Computing},
  volume={45},
  number={4},
  pages={A2101--A2118},
  year={2023},
  publisher={SIAM}
}

@article{monturiol2024tensorkrowch,
  title={TensorKrowch: Smooth integration of tensor networks in machine learning},
  author={Monturiol, Jos{\'e} Ram{\'o}n Pareja and P{\'e}rez-Garc{\'\i}a, David and Pozas-Kerstjens, Alejandro},
  journal={Quantum},
  volume={8},
  pages={1364},
  year={2024},
  publisher={Verein zur F{\"o}rderung des Open Access Publizierens in den Quantenwissenschaften}
}

@article{sehlstedt2025software,
  title={The Software Landscape for the Density Matrix Renormalization Group},
  author={Sehlstedt, Per and Brandejs, Jan and Bientinesi, Paolo and Karlsson, Lars},
  journal={arXiv preprint arXiv:2506.12629},
  year={2025}
}







\end{document}